\documentclass[conference,compsoc]{IEEEtran}

\IEEEoverridecommandlockouts 

\PassOptionsToPackage{dvipsnames,table}{xcolor}
\usepackage{graphicx}
\usepackage{amsmath}
\usepackage{amsfonts}
\usepackage{graphicx}
\usepackage{listings}
\usepackage{hyperref}
\usepackage{soul}
\usepackage{booktabs}
\usepackage{enumitem}
\usepackage{tcolorbox}
\usepackage{subfig}
\usepackage{verbatim}
\usepackage{fontawesome5}
\usepackage{pifont}
\usepackage{array}

\ifCLASSOPTIONcompsoc
  \usepackage[nocompress]{cite}
\else
  \usepackage{cite}
\fi

\makeatletter
\newcommand{\linebreakand}{%
  \end{@IEEEauthorhalign}
  \hfill\mbox{}\par
  \mbox{}\hfill\begin{@IEEEauthorhalign}
}
\makeatother

\title{LLMs in the SOC: An Empirical Study of Human-AI Collaboration in Security Operations Centres}

\author{
\IEEEauthorblockN{Ronal Singh}
\IEEEauthorblockA{\textit{Data61, CSIRO} \\
Melbourne, Australia}
\and
\IEEEauthorblockN{Shahroz Tariq\IEEEauthorrefmark{1}\thanks{\IEEEauthorrefmark{1}Authors contributed equally as second authors.}}
\IEEEauthorblockA{\textit{Data61, CSIRO} \\
Sydney, Australia}
\and
\IEEEauthorblockN{Fatemeh Jalalvand\IEEEauthorrefmark{1}}
\IEEEauthorblockA{\textit{Data61, CSIRO} \\
Melbourne, Australia}
\and
\IEEEauthorblockN{Mohan Baruwal Chhetri}
\IEEEauthorblockA{\textit{Data61, CSIRO} \\
Melbourne, Australia}

\linebreakand

\IEEEauthorblockN{Surya Nepal}
\IEEEauthorblockA{\textit{Data61, CSIRO} \\
Sydney, Australia}
\and
\IEEEauthorblockN{Cecile Paris}
\IEEEauthorblockA{\textit{Data61, CSIRO} \\
Sydney, Australia}
\and
\IEEEauthorblockN{Martin Lochner}
\IEEEauthorblockA{\textit{eSentire Inc.} \\
Waterloo, Canada}
}
\date{September 2025}

\begin{document}

\maketitle

\begin{abstract}

The integration of Large Language Models (LLMs) into Security Operations Centres (SOCs) presents a transformative, yet still evolving, opportunity to reduce analyst workload through human-AI collaboration. However, their real-world application in SOCs remains underexplored. %
To address this gap, we present a longitudinal study of 3,090 analyst queries from 45 SOC analysts over 10 months. 
Our analysis reveals that analysts use LLMs as on-demand aids for sensemaking and context-building, rather than for making high-stakes determinations, preserving analyst decision authority. 
The majority of queries are related to interpreting low-level telemetry (e.g., commands) and refining technical communication through short (1-3 turn) interactions.
Notably, 93\% of queries align with established cybersecurity competencies (NICE Framework), underscoring the relevance of LLM use for SOC-related tasks. 
Despite variations in tasks and engagement, usage trends indicate a shift from occasional exploration to routine integration, with growing adoption and sustained use among a subset of analysts. 
We find that LLMs function as flexible, on-demand cognitive aids that augment, rather than replace, SOC expertise. 
Our study provides actionable guidance for designing context-aware, human-centred AI assistance in security operations, highlighting the need for further in-the-wild research on real-world analyst-LLM collaboration, challenges, and impacts.

\end{abstract}

\section{Introduction}
\label{sec:intro}

\begin{figure}[t]
    \centering
    \includegraphics[width=1\linewidth]{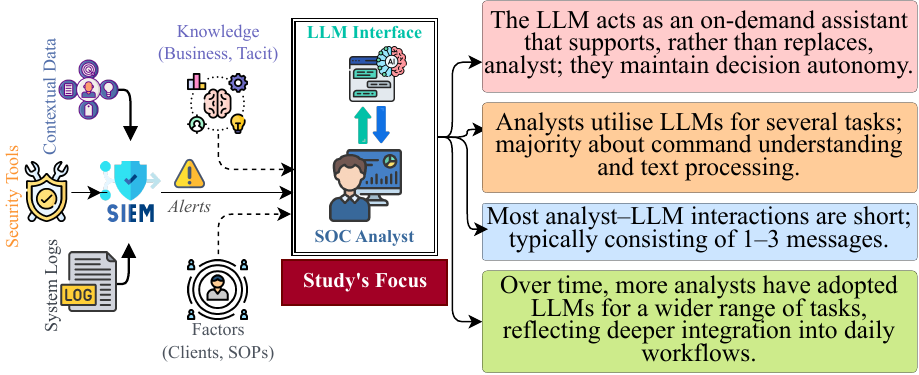}
    \caption{SOC workflow, focus of study and our insights.}
    \label{fig:interface}
\end{figure}

SOCs handle diverse functions, from real-time detection and incident response to continuous improvement~\cite{vielberth2020security,Baruwal-Chhetri2024-yf,SOC_10Strategies}. SOC analysts play a critical role in this by interpreting data and making informed decisions. They typically follow a multi-stage process, beginning with quick assessments of alert relevance~\cite{jalalvand2024alert}, then checking for related signals, gathering contextual telemetry to build situational awareness~\cite{SOC_10Strategies}, and evaluating evidence of compromise to make escalation decisions~\cite{Kersten2023-wy}. This reasoning workflow is supported by tooling like SIEM, SOAR, and XDR~\cite{SOC_10Strategies,tariq2025alert}.

Recently, LLMs have emerged as promising tools to support such SOC workflows, with applications ranging from threat intelligence to triage automation~\cite{Freitas2024-sx,Siracusano2023-qs,Sufi2024-vc, Feffer2024-wt}. While prior research explores human-LLM interaction in various domains and lab studies~\cite{gao2024taxonomy,kim2025applying,guo2024investigating,Ullah2024-tl}, it does not capture how SOC analysts utilise LLMs in their daily workflows. Existing work, such as Copilot Guided Response (CGR)~\cite{Freitas2024-sx}, shows system-level performance, not individual analysts' usage patterns.

Gaining insights into how SOC analysts utilise LLMs in practice, including the types of questions they ask and the tasks they use them for, is essential for evaluating the practical role of LLMs in operational cybersecurity (Figure~\ref{fig:interface}). This study addresses that gap by analysing 3,090 queries submitted to GPT-4 during live investigations over 10 months (May 2023 to March 2024) by 45 SOC analysts at eSentire Inc., a cybersecurity firm specialising in 24/7 managed detection and response \cite{Lochner2025-zc}. To guide this investigation, we focus on the following research question:

{
\begin{tcolorbox}
[width=1\linewidth, center,  left=5pt, right=5pt, top=2pt, bottom=2pt,label=rq1,colback=Cerulean!15,colframe=Cerulean!15,boxrule=0.5pt]

\textit{How do SOC analysts utilise LLMs in their daily workflows, including the specific tasks they apply them to, how these uses align with established cybersecurity frameworks (e.g., NICE), and the conversational patterns that characterise their interactions?
}
   
\end{tcolorbox}
}

The answers to this question reveal the task priorities and engagement patterns of SOC analysts with LLMs, informing the design of more effective LLM-based assistants as human-AI collaboration becomes increasingly common in SOC environments. We address this question through a multi-level analysis of analyst-LLM interactions across queries, tasks, and conversations.

\begin{itemize}
    \item 
        \textbf{Query-level:} We examine \textit{what} analysts ask and \textit{how} they phrase their requests, revealing their cognitive needs, the daily tasks augmented by LLMs, and alignment with frameworks such as NICE.
    \item 
        \textbf{Conversation-level:} We analyse sequences of queries within investigative sessions to uncover \textit{how} analysts structure multi-turn interactions in their daily workflows. This reveals reasoning patterns and functional roles that LLMs assume in conversations.
    \item 
        \textbf{Analyst-level:} We investigate the frequency and manner in which different analysts use LLMs. This variation reveals diverse strategies for integrating LLMs and highlights how they become embedded in the daily workflows of different analysts.
\end{itemize}

We began with descriptive analysis to map overall query activity, including how frequently analysts interacted with the LLM. Then, we conducted a thematic analysis~\cite{Nowell2017-ud} to identify patterns in query phrasing, topics, and tasks. Since our research objective was to characterise analyst behaviour, not evaluate LLM performance, we focused our thematic analysis on analyst queries and used LLM responses only to contextualise intent. We grouped successive analyst queries on the same topic into single conversational units, based on topic continuity. Finally, we synthesised insights across  Query, Conversation, and Analyst levels to characterise analyst-LLM interactions. Although our dataset covers a single enterprise SOC using one type of text-based LLM, it provides a valuable longitudinal view of real-world LLM use, highlighting key usage patterns and tasks. As more studies emerge, both quantitative (e.g., \cite{Freitas2024-sx}) and qualitative (e.g., ours, \cite{Mink2023-ea,Kersten2023-wy}), research and design communities will be better informed to design targeted LLM-based SOC tools.

\subsubsection*{Contributions}

We make the following contributions:

\begin{itemize}
    \item  
    We present the \textbf{first in-depth, real-world analysis of LLM adoption in a SOC}, spanning 3,090 queries from 45 analysts over 10 months. Our analysis highlights analyst-level usage patterns across task types and individual differences.

    \item 
    We develop a \textbf{structured coding framework that facilitates thematic analysis of analyst-LLM interactions} by classifying LLM queries according to task type, subject, and interaction pattern. We apply this framework to develop a deeper understanding of how LLMs assist investigative, explanatory, and communicative functions within SOC workflows.

    \item We provide \textbf{empirical insights into analyst behaviour and LLM utility in SOCs.} We identify three core use cases that characterise how SOC analysts engage with LLMs: (i) interpreting raw telemetry, (ii) generating and refining task-related communication content, and (iii) obtaining rapid, on-demand assistance through concise interactions. Collectively, our analysis reveals that LLMs are becoming routine, competency-aligned aides rather than decision-makers: 93\% of queries map to NICE tasks, usage has shifted from ad-hoc exploration to sustained integration, and analysts retain final judgement while leveraging LLMs for rapid sense-making. These findings guide the design of context-aware, human-centred assistants to enhance human-AI collaboration in SOCs.
    
    \item We offer \textbf{design recommendations} for AI-augmented SOCs: (1) embed analyst- and context-adaptive LLM explanations of technical artifacts to speed triage and support evidence-based decisions; (2) integrate LLM assistants for microtasks to provide in-workflow support and minimise context switching; and (3) surface evidence over recommendations for investigative tasks, preserving analyst autonomy and context.
    
\end{itemize}

\section{Related Works}
\label{relatedworks}

\textbf{AI for SOCs.} SOCs integrate multiple systems including SIEM platforms that ingest telemetry from IDS/IPS, firewalls,  and EDR tools; SOAR platforms for response coordination; and supplementary tools such as User and Entity Behaviour Analytics (UEBA) and Threat Intelligence Platforms (TIPs) \cite{Binbeshr2025-hi,Khayat2025-qx}. These systems have long relied on AI to enhance operational efficiency and decision-making. Prior to the emergence of LLMs, AI was already embedded across the SOC workflow, from threat detection \cite{Yang2024-vo}, alert triage \cite{Chavali2024-ly}, and incident response \cite{Sworna2023-jm} to end-to-end automation \cite{Khayat2025-qx}. SIEMs employed ML/DL for event correlation, anomaly detection, and prioritisation \cite{Binbeshr2025-hi}, while TIPs used NLP and graph inference to enrich contextual understanding \cite{Khayat2025-qx}. Commercial tools such as Microsoft Sentinel, Maltego, and Splunk exemplify this integration, offering AI-driven capabilities for visualisation, triage, and orchestration \cite{Freitas2024-sx,Binbeshr2025-hi,Khayat2025-qx,Hasanov2024-hw,tariq2025alert}. While effective, these approaches were task-specific and limited in their ability to reason across diverse, unstructured contexts. The recent emergence of LLMs provides a broader, context-aware form of cybersecurity reasoning, enabling more adaptive and human-aligned support across SOC operations \cite{Hasanov2024-hw}.

\textbf{LLMs for Cybersecurity.}  
Broadly, LLMs have been used to support vulnerability scanning, anomaly/intrusion detection, phishing simulation, threat intel summarisation, and privacy-preserving analysis~\cite{Deng2024-jk,Ferrag2023-bu,Thomas2024-dt,Manocchio2024-cl,Hassanin2025-eo,tariq2025a2c,Koide2024-jx,tariq2025bridging,cuong2025towards,Mitra2024-cg}. Recent works highlight the expanding role of LLMs in cybersecurity workflows to support SOC analysts. For example, Copilot Guided Response (CGR) aids triage and remediation by generating recommendations from historical threats~\cite{Freitas2024-sx}. Most of these systems are evaluated in controlled environments, with a focus on performance metrics such as speed and accuracy of decision-making. In contrast, our study offers the first empirical analysis of LLM use in a live SOC, examining how analysts engage with LLMs, what help they seek, how interactions unfold, and how adoption evolves, providing a unique human-centred view of analyst-LLM practice. 

\textbf{LLM Roles in Decision Support.}
LLMs play diverse roles in human-AI decision-making, shaped by system autonomy, user needs, and interaction structure~\cite{Jiang2024-eq, Song2024-mw, Eigner2024-zd, Reicherts2025-we, Ma2024-uh}. LLMs span roles from \textit{assisted intelligence} (retrieval and summarisation) to more collaborative forms like \textit{augmented} and \textit{cooperative intelligence}, which aid reasoning and joint problem solving~\cite{Jiang2024-eq, Reicherts2025-we}. HCI research stresses aligning AI support with decision phases, whether recommending actions, predicting, proposing alternatives, or surfacing evidence~\cite{Pang2025-vk, Eigner2024-zd, Song2024-mw, Reicherts2025-we}. While AI suggestions can speed decisions, they may also increase over-reliance~\cite{Fogliato2022-dj, Bucinca2021-di}. Cognitive forcing techniques (e.g., delaying AI output) can reduce over-reliance but risk under-reliance~\cite{Fogliato2022-dj, Gajos2022-wh}. \textit{Evaluative AI} provides balanced insights, highlighting pros and cons without dictating choices, thus supporting autonomy in machine-in-the-loop systems~\cite{Miller2023-uz,Le2024-mc,Shneiderman2022-uj,Singh2025-gy}. We examine how LLMs function in practice and what \textit{decision support} analysts seek.

\textbf{Interaction Patterns and Analyst Strategies.}

Despite the growing capabilities of models, interaction patterns between humans and LLMs remain underexplored. Studies show that interaction framing and modality shape user expectations, trust, and behaviour~\cite{guo2024investigating, li2024map, kim2025applying, gao2024taxonomy, chan2024human, schneider2024exploring, tariq2025bridging}. Recent work highlights that prompt structure shapes explainability and collaboration~\cite{guo2024investigating}, dialogue in complex tasks involves verbal and nonverbal cues~\cite{chan2024human}, and human--AI interaction spans planning to testing across varying agent roles~\cite{gao2024taxonomy}. These studies show that effective LLM integration depends on their ability to adapt to user needs, interaction design, and task context. Our study addresses this gap by analysing thousands of analyst-LLM interactions in an operational SOC, uncovering concrete interaction patterns, evolving strategies, and the situated use of LLMs as cognitive aids in high-stakes, time-sensitive environments.

\section{Method}
\begin{figure}[t!]
    \centering
    \includegraphics[trim=0pt 0pt 0pt 0pt, clip,width=0.95\linewidth]{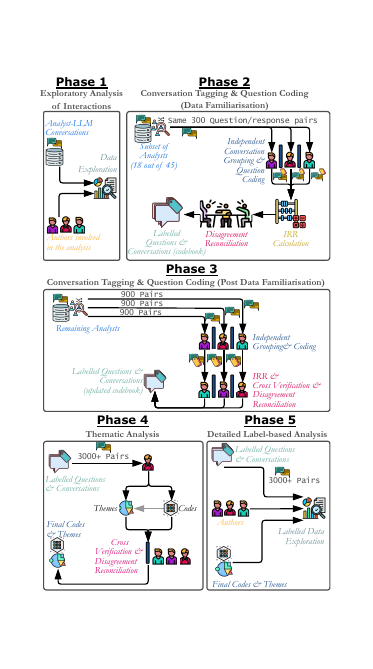}
    \caption{The five-phased approach to analyse and understand SOC analysts' interactions with LLMs.}
    \label{fig:SOC_LLM_Method_v2}
\end{figure}

To analyse and understand SOC analysts' interactions with LLMs, we adopted a five-phase methodology, as illustrated in Figure~\ref{fig:SOC_LLM_Method_v2}, which we describe in detail below. This approach is guided by Nowell et al.'s~\cite{Nowell2017-ud} multi-phase framework for thematic analysis, encompassing: (i) familiarisation with data (phases 1-2), (ii) code generation (phases 2-3), (iii) theme identification, review, and definition (phase 4), and (iv) reporting (phase 5). We follow their iterative process and incorporate specific trustworthiness practices, including multi-coder independent coding with strong inter-rater reliability (IRR), triangulation, reconciliation, cross-verification, and systematic documentation.  

\vspace{5pt}
\noindent
\textbf{\textsc{Phase 1}: Exploratory Analysis of Interactions.} In this phase, we performed a statistical analysis of analyst-LLM interactions to characterise overall usage patterns, such as the total number of queries, engagement frequency, and query length distributions, providing a quantitative foundation for understanding analyst behaviours before moving to deeper thematic and qualitative analysis.

\vspace{5pt}
\noindent
\textbf{\textsc{Phase 2}: Data Familiarisation for Question Coding \& Conversation Tagging (Partial Dataset).} This phase involved three members of the research team reviewing the question-LLM response pairs multiple times to develop a shared understanding of the analyst queries as well as identify and tag conversations for thematic analysis. Since the data consisted of individual analyst-LLM interactions, the team first needed to group these into coherent conversations.

\vspace{5pt}
\noindent
\textit{\textbf{Question Coding.}} 
This step focused on developing a coding scheme to analyse the nature of the analysts' queries. The three researchers first familiarised themselves with the data by reviewing a range of question-LLM response pairs. Through discussion, the team agreed on three key attributes to enable a meaningful decomposition of the queries:

\begin{itemize}
    \item 
        \textbf{Query Pattern:} The structural form or phrasing of the analyst's input, abstracted to remove specific details (e.g.,  filenames, commands, URLs). For example, a pattern such as \textit{``What is this [$\ldots$] doing''}, captures a common syntactic form regardless of the specific artifact used. This abstraction facilitates grouping similar queries and supports broader pattern analysis.
        
    \item 
        \textbf{Query Subject:} The primary topic or focus of the question, indicating the subject matter the analyst is asking about (e.g., a command, malware, or file).  
    \item 
        \textbf{Task:} The underlying intent behind the query, categorised into functional areas such as \textit{Command Understanding} or \textit{Summarising LLM Output}.
\end{itemize}

To ensure consistent application of the agreed coding scheme, all three researchers independently coded a sample of 20 question-LLM response pairs. This sample was used to verify their initial shared understanding of coding the three attributes. This was followed by a calibration session, during which the same 20 pairs were jointly reviewed and coded to reconcile differences, refine the definitions, and develop a preliminary codebook. Once alignment was achieved, the researchers independently coded a further 310 (10\%) question-response pairs to further test their collective understanding of the attributes and coding scheme before being confident in independently coding the remaining pairs. IRR, measured using Fleiss' Kappa across the three dimensions, showed strong agreement: 0.90 for Query Pattern, 0.82 for Query Subject, and 0.79 for Task, indicating substantial to near-perfect agreement. A reconciliation session helped resolve any disagreements and update the codebook. 

\vspace{5pt}
\noindent
\textit{\textbf{Conversation tagging.}} Following a similar process described above, the next step involved grouping successive analyst queries that relate to the same topic or investigative thread and treating them as a single conversational unit by considering (a) overlapping security artifacts (e.g., commands, email rules), (b) continuity in investigative intent (e.g., both steps probing the same log entry), and (c) occurring within a 30-minute window (used only as a guideline rather than a strict cutoff). Each conversation was assigned a unique identifier (e.g., \textit{C1}, \textit{C2}). To ensure consistency and reliability in tagging, the three researchers independently labelled a random sample of 45 analyst questions, comprising 142 question–response pairs, then met to reconcile and refine their understanding of the tagging process. Once consistent tagging practices were established, an additional 240 instances (13\%) were tagged as a further verification step. Fleiss' Kappa was used to calculate IRR, resulting in a score of 0.75, indicating substantial agreement. Differences were reconciled through discussions.

\vspace{5pt}
\noindent
\textbf{\textsc{Phase 3}}: Remaining Dataset Question Coding \& Conversation Tagging. In this phase, given acceptable agreement in Phase 2, the remaining question-LLM response pairs were divided among the three researchers, with each coding approximately 900 pairs independently. After initial coding, each researcher's annotations were cross-verified by another team member to ensure consistency and accuracy; discrepancies were resolved through discussions. To further enhance the reliability of the coding process, the lead author conducted a comprehensive review of all the coded data. Discrepancies were minimal, observed only in 4\% of the cases and resolved collaboratively through discussions. 

\vspace{3pt}
\noindent
\textbf{\textsc{Phase 4}: Query Level Thematic Analysis. }
One researcher consolidated all analyst question-LLM response pairs into a dataset and conducted query-level thematic analysis~\cite{Clarke2017-au}, as outlined in Figure~\ref{fig:SOC_LLM_Method_v2}. This phase built on earlier manual coding (\textsc{Phase 3}) and introduced a hybrid approach that combined qualitative coding with data-driven clustering and thematic interpretation (\textsc{Phase 5}) to surface higher-level insights. We applied semantic clustering to the coded data to support theme development. Sentence-BERT embeddings (\texttt{stsb-roberta-large}) and agglomerative clustering with cosine similarity~\cite{Abdalgader2024-fg} were used to group over 1000 distinct query patterns into semantically similar codes, supporting pattern abstraction. All clusters were manually reviewed to ensure semantic coherence, considering the original query and LLM response. Similar patterns such as \textit{explain [x]} and \textit{explain this [x]} were grouped. Themes were then developed by merging related clusters and assigning meaningful, high-level labels spanning across all attributes; e.g., \textit{command} and \textit{multiple commands} formed a broader \textit{command-related subject} theme. For example, semantic clustering was applied to cluster 1,354 initial query pattern codes into 980 unnamed clusters, which were iteratively merged by the lead author into 147 codes and finally grouped into themes.

To ensure the robustness of these themes, we conducted a \textit{triangulation process}~\cite{Nowell2017-ud} in which two researchers, who were not involved in the clustering, independently reviewed and refined the themes. This hybrid approach, combining coding, clustering, and human insight, balanced rigour with interpretive depth.~\cite{Louwers2024-za, Chen2024-am}.

\vspace{5pt}
\noindent
\textbf{\textsc{Phase 5}: Conversation and Analyst Level Thematic Analysis.}
In the final phase, we conducted an in-depth analysis using the final annotated dataset from \textsc{Phase 4}. This involved both conversation-level and analyst-level analysis to uncover patterns at both levels. We first analysed the structure, frequency and patterns of conversations, identifying common interaction styles and task transitions. Analysts were then clustered based on usage volume, revealing distinct engagement profiles and thematic preferences. Finally, we explored behavioural and cognitive patterns, including low-engagement users, to understand diverse styles of LLM integration and potential disengagement plans.

\section{Results}
\label{sec:results}

\begin{table*}[ht!]
\centering
\begin{tabular}{@{} l c p{13cm} @{}}
\toprule
\textbf{Analyst Role} & \textbf{Count} & \textbf{Responsibilities} \\
\midrule
SOC Analyst I & 24 & Handles initial alert triage and prioritisation, basic malware and network traffic analysis, IP/domain reputation checks, OSINT research, incident documentation, false positive tuning, and escalates complex cases.
 \\
SOC Analyst II & 4 & Coordinates intermediate incident response, manages specialised security tools, performs advanced log analysis and digital forensics, develops automation scripts, liaises with vendors, and reviews junior analysts' work. \\
Senior SOC Analyst & 7 & Leads advanced investigations and root cause analysis, conducts malware reverse engineering, runs threat hunting campaigns, mentors staff, optimises SIEM rules, coordinates across teams, and drives process improvements. \\
SOC Incident Handler & 1 & Collects and analyses threat intelligence, develops IOCs, attributes threat actors, monitors threat landscape, produces reports for leadership, integrates external feeds, and performs strategic risk assessments. \\
Threat Analyst & 6 & Oversees major incident response, containment, and eradication, handles stakeholder communication and coordination, facilitates post-incident reviews, executes emergency procedures, and supports continuity planning.  \\
Associate Analyst & 3 & Provides entry-level monitoring, basic research, alert categorisation, report generation, and administrative support, while engaging in training and certification programs. \\
\bottomrule
\end{tabular}
\caption{Roles and responsibilities of the 45 participating SOC analysts.}
\label{tab:soc_analysts_demo}
\end{table*}

\begin{figure*}[!ht]
    \centering
    \includegraphics[width=\linewidth]{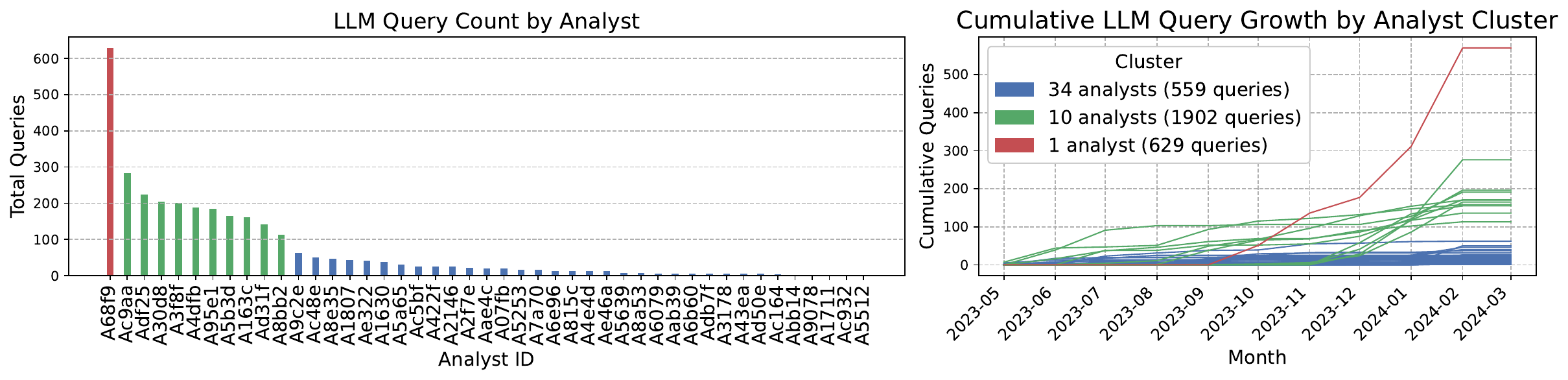}
    \caption{(Left) Query volumes vary across analysts, with heavy concentration among a few users. (Right) Overall, there is growing integration into workflows, but mostly driven by a subset of analysts (March 2024 has only 7 days of data).}
    \label{fig:llm_analyst_engagement_summary}
\end{figure*}

This section presents findings from our analysis of LLM usage. We start with an overview of SOC analysts, model information, and the dataset before diving into exploratory analysis. We then present the findings from a three-level thematic analysis covering query, conversation, and analyst levels. The dataset comprised 3,122 analyst-submitted queries to GPT-4, collected between May 18, 2023, and March 7, 2024. After removing 32 queries where the LLM could not provide a meaningful response (e.g., due to missing input or unsupported tasks), 3,090 valid queries and 532 distinct conversations remained for analysis. 

\subsection{Participant Overview and LLM Usage Constraints} 

\textbf{Participants.} Table~\ref{tab:soc_analysts_demo} summarises the roles and responsibilities of the 45 SOC analysts who participated in the study. The majority were Tier 1 (SOC Analyst I), reflecting their front-line role in alert triage. Tier 2 (SOC Analyst II) and Tier 3 (Threat Analyst, Incident Handler) roles are more specialised and fewer in number, accounting for their lower representation. All participants, except for Associate Analysts (interns/co-ops, n=3), had 3--5 years of experience. Participation in the study was voluntary and the distribution of Tier 1, Tier 2, and Tier 3 analysts is not representative of actual staffing ratios. Analysts opted into the LLM pilot study via an internal Slack announcement, which invited exploratory use of the tool for work-related tasks. 

\textbf{LLM Setup.} Participants accessed GPT-4 (GPT-4-0613) via a browser-based interface to an internally hosted ``LLM-Gateway'',\footnote{\url{https://github.com/eSentire-Labs/LLM-Gateway}}, which provided a control and monitoring layer for LLM usage. The gateway enabled API-based access to OpenAI's GPT-4 Enterprise platform under a 30-day data deletion policy and a non-training agreement in place. Analysts were free to navigate to the system and use the model to support their work tasks. The model was used without fine-tuning and had no access to the internet or internal SOC systems. Although other models were available, this study focuses exclusively on GPT-4.

\textbf{Data Sources.} Analysts used proprietary and third-party tools to access telemetry from endpoints, networks, identities, and application layers, including EDR, NDR, SIEM/XDR, firewalls, DNS, email, identity, cloud, WAF/CDN, and SaaS applications. Participants were permitted to use only Public and Internal data, with strict policies prohibiting submission of Confidential or Highly Confidential information\footnote{Data was classified into four categories based on organisational sensitivity: Public data is freely accessible; Internal data comprises non-sensitive organisational information; Confidential data is protected by legal or contractual obligations; and Highly Confidential data is subject to the most stringent controls}. Aside from this restriction, analysts were free to engage with the LLM as they saw fit. As the study focused on initial LLM engagement, all data were anonymised and devoid of personal identifiers. Accordingly, we report only aggregate usage trends, without attributing interactions to individual roles or experience levels.

\subsection{Exploratory Analysis}
\label{sec:results-exploratory-phaseI}

Figure~\ref{fig:llm_analyst_engagement_summary} (left) summarises analyst engagement with the LLM over 10 months. Usage was uneven, with a small number of analysts generating the majority of queries.

Given the skewed usage, we clustered the analysts based on their usage before analysing their long-term reliance on the LLM. We applied the Fisher-Jenks natural breaks algorithm\footnote{\url{https://en.wikipedia.org/wiki/Jenks_natural_breaks_optimization}} to total query counts per analyst. This generated three distinct clusters: a large group of \textit{low-usage analysts} (Cluster 0, $n=34$), a smaller group of \textit{moderate users} (Cluster 1, $n=10$), and a single \textit{power user} (Cluster 2, $n=1$; most active analyst). Figure~\ref{fig:llm_analyst_engagement_summary} (right) shows that daily usage increased steadily over time, driven by a subset of highly active analysts (a stronger increase was observed among 11 analysts), suggesting deeper integration of the LLM into daily workflows. The number of active analysts also grew month on month, suggesting sustained interest and continuous adoption of the LLM. 

While a single power user submitted approximately 600 queries (17\% of the total dataset), usage was not limited to a few individuals. By early 2024, all 45 analysts had interacted with the system, and daily query volumes increased from fewer than 10 queries per day in May-August 2023 to over 30 queries per day in January and February 2024. We analysed the monthly usage patterns and observed that the number of active analysts grew from 4 in May 2023 to 13 in June, reaching 18 by July and steadily increasing to a peak of 25 in February 2024. By early March 2024, 16 analysts had already submitted queries, showing strong ongoing engagement.

The monthly increases include the proportion of returning analysts and new analysts rising each month (see Figure~\ref{fig:analyst_engagement_metrics} in the Appendix). By early 2024, the majority of LLM interactions came from returning analysts, indicating ongoing integration of the tool into daily workflows. Most active analysts utilise the LLM in tightly clustered bursts, with a median gap of 1-2 hours between visits, suggesting its use as an on-demand aid between investigative steps.

The average query length was 25 words (standard deviation: 70), but this distribution was skewed by a few very long queries. Some analysts submitted highly verbose queries, exceeding 1000 words (due to code or script segments), while others consistently relied on concise prompts of fewer than 20 words. In contrast, LLM responses were more uniformly verbose, with a mean of 161 words (standard deviation: 97) and a relatively narrow interquartile range.

\subsection{Query Level Analysis}
\label{sec:results-thematic-phaseIV}

This section discusses the thematic analysis results of \textit{individual queries}. Each \textit{conversation} could involve multi-step interactions. The thematic analysis was conducted on individual queries, rather than entire conversations, to retain fine-grained insights that a holistic approach might overlook. Section~\ref{sec:results-conversation-phaseV} presents conversation-level analysis. \textbf{The appendix includes the list of all themes; full codebook can be accessed via the link in the footnote:}\footnote{\url{https://osf.io/g8tr6/?view_only=db9c65552d364cba8c3a499bd12df589}}.

Analysts varied widely in usage patterns, with some submitting over 600 queries and others fewer than 5. To assess the robustness of our results to outlier effects following thematic analysis, we recalculated the proportion of queries assigned to each task theme after excluding (i) the most active analyst, (ii) the top 10 most active analysts, and (iii) both the most active and the 20 least active analysts. Across all scenarios, the rank order of the top ten and the top two themes remained unchanged. 

{
\begin{tcolorbox}
[width=1\linewidth, center,  left=10pt, right=10pt, top=2pt, bottom=2pt,label=res1,colback=Orchid!10,colframe=Orchid!10,boxrule=0.5pt]
   The findings from the exploratory analysis reveal an interesting landscape of LLM use in our dataset. Despite wide variation in query volume, engagement style, and query lengths, \textbf{usage trends point to} a transition from periodic exploration to routine integration, with \textbf{increased adoption over time and sustained interaction by a subset of analysts, with a stronger adoption among 11 analysts}. These observations offer grounded insights into how LLMs are appropriated in practice, setting the stage for the next phase of our study: a qualitative analysis of the tasks and interaction strategies.
\end{tcolorbox}
}

\begin{figure}[!t]
    \centering
    \includegraphics[width=0.99\linewidth]{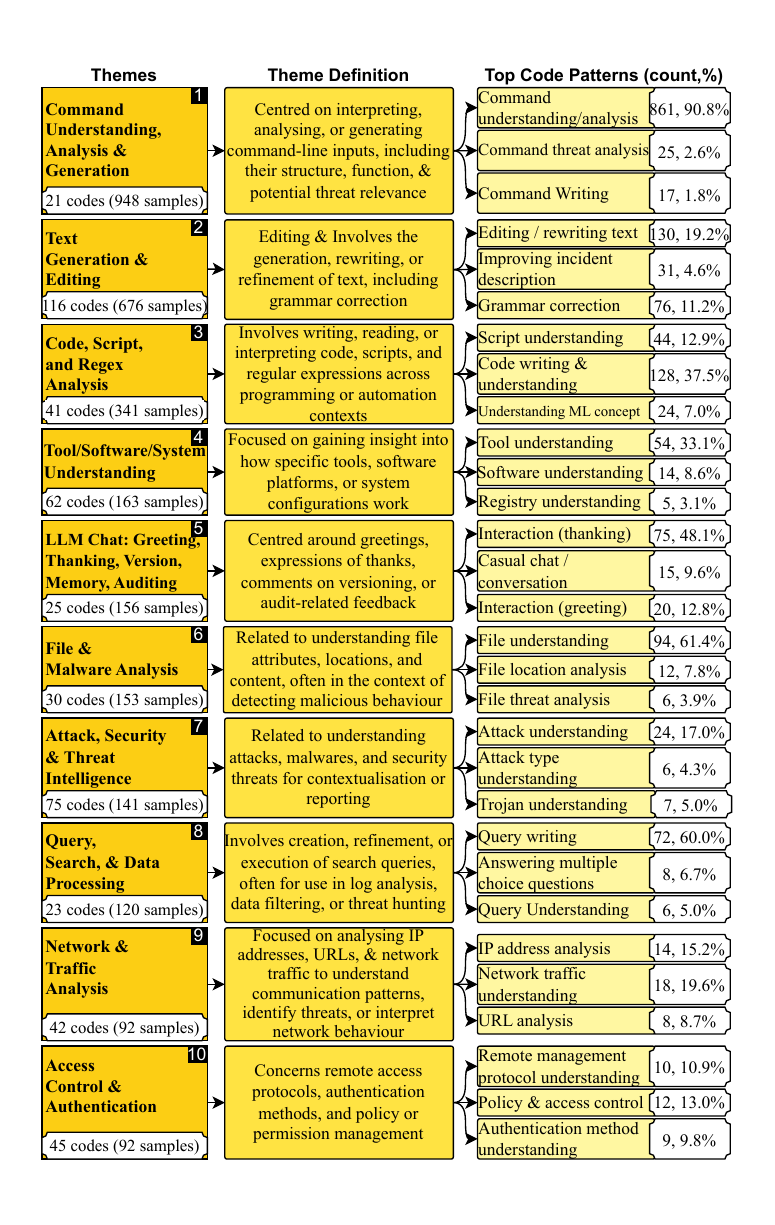}

    \caption{Summary of Task Themes (top 10)}
    \label{fig:task-themes}
\end{figure}

\subsubsection{SOC Analyst Task Types}

Task themes reflect the actual work that analysts aim to accomplish when interacting with an LLM. \textbf{We find that analysts used LLMs for a range of tasks central to day-to-day SOC operations}~\cite{vielberth2020security,SOC_10Strategies}. These include interpreting and analysing commands; editing and improving the clarity of incident descriptions, documentation, and alert text; and understanding or troubleshooting code, scripts, and regular expressions. Analysts also used LLMs to analyse malware, explore system behaviours, build queries, and clarify email rules, highlighting the breadth of operational tasks. We highlight three common task themes below (see Figure~\ref{fig:task-themes} for the top 10), showing how LLMs integrate into workflows. For task adoption over time, see Figure~\ref{fig:month_theme_balloon_plot} in the appendix. 

\subsubsection*{Command Understanding, Analysis \& Generation} This was the most frequent task theme (31\% of queries). Queries typically involved analysts submitting command-line instructions and requesting clarification on functionality, syntax, or behavioural purpose. Common prompts such as \textit{``what does [x] do''} or \textit{``explain this [x]''} ([x] is the actual command(s)) suggest investigative or verification-oriented goals, with the LLM supporting the analyst through interpretive scaffolding. While we did not perform a thorough thematic coding of the LLM responses, we \textbf{did} analyse the responses it provided for the command-related questions to understand why this theme was likely prominent.

The LLM often began with an overview of the command, explaining the components and their relationships, inferring the intent behind complex operations, and highlighting suspicious behaviours. For example: \textit{``In summary, the [anonymised command] is trying to login as [anonymised user] without password and $\ldots$''}.

Many responses also included parameter breakdowns, example outcomes, risk implications, and recommended follow-up actions. Based on their sustained usage, \textbf{we hypothesise that by decomposing command components and synthesising likely intent and behaviour, the LLM has shown the potential of enhancing the analysts' situational understanding}. This aligns with prior work (e.g.,~\cite{Mink2023-ea}), which finds that SOC analysts prefer explanations that contextualise evidence and inform action.

\subsubsection*{Text Generation \& Editing} This theme, comprising 22\% of queries, included requests to rephrase incident descriptions or alert summaries or correct grammar. Analysts frequently used queries such as \textit{``make this better''}, \textit{``can you correct this''}, or \textit{``rephrase''}. \textbf{This theme underscores the increasing expectation that LLMs will assist SOC analysts in producing clear, coherent written documentation for SOC systems and peer or client communication}.

\subsubsection*{Code, Script, \& Regex Analysis} Representing 11\% of queries, this theme included submissions of code blocks (e.g., Python, PowerShell, JavaScript) and regex expressions. Analysts requested help understanding their behaviour, decoding strings, or generating new logic. Example queries included \textit{``what does this script do''}, or \textit{``write a regex to detect [x]''}. This theme suggests that \textbf{analysts and the LLM engaged in complex reasoning around syntax, logic, and security behaviour of scripts and code}. Regex generation was also notable, typically used in filtering logs.

\subsubsection*{Other Technical Tasks} In addition to the top three themes, analysts used LLMs for a wide range of other tasks including \textit{Tool/ Software/ System Understanding}; \textit{Attack, Security \& Threat Intelligence}; \textit{Access Control \& Authentication}; \textit{File \& Malware Analysis}; \textit{Network \& Traffic Analysis}; and \textit{Query, Search \& Data Processing}. These tasks highlight specialised needs where LLMs offer targeted support. Notably, \textit{LLM Chat} interactions reveal a growing rapport, with five analysts explicitly expressing thanks.

{
\begin{tcolorbox}
[width=1\linewidth, center,  left=10pt, right=10pt, top=2pt, bottom=2pt,label=res2,colback=Orchid!10,colframe=Orchid!10,boxrule=0.5pt]
   The results reveal a \textbf{growing reliance on LLMs across multiple functions:} from interpreting low-level artifacts, to generating or refining work products such as incident notes, queries, and email, as well as conducting on-demand look-ups, including definitions. \textbf{Findings suggest that LLMs are emerging as flexible \textit{cognitive aids}} (e.g., as explainer or interpreter, drafting assistant, coding helper, and on-demand reference tool) rather than just static chatbots. The diversity of tasks calls for a unified integration.
\end{tcolorbox}
}

\subsubsection{SOC Analyst Query Subjects}

We encoded query subjects to preserve anonymity (to not disclose proprietary information), resulting in a strong one-to-one mapping between subjects and task themes. For example, the task \textit{Command Understanding} aligns with the subject \textit{command}. 

Across the 3,090 coded queries, subject matter clustered into a few dominant types: \textit{command}-related artifacts (e.g., \textit{powershell} and \textit{curl} commands) led by a wide margin ($\approx$30\%), followed by requests about phrasing or formatting snippets of written content ($\approx$17\%) and code or scripting issues such as regex or PowerShell ($\approx$9\%). Tool-specific questions ($\approx$7\%) and file- or directory-path look-ups ($\approx$5\%) rounded out the top five. The remainder was dispersed across LLM meta-chat, malware or threat look-ups, MDR/EDR query tuning, alert narration, network indicators, email-rule checks, and assorted `how-to' prompts.

{
\begin{tcolorbox}
[width=1\linewidth, center,  left=10pt, right=10pt, top=2pt, bottom=2pt,label=res3,colback=Orchid!10,colframe=Orchid!10,boxrule=0.5pt]
   \textit{\textbf{Telemetry tells us what was observed; analyst queries reveal which aspects they seek to interpret using LLMs}}. While security teams collect rich telemetry via tools like SIEMs and EDRs, LLMs are increasingly used to interpret parts of that raw input, transforming it into potentially actionable insights or contextual understanding. Analysts appear to have relied on LLMs to interpret telemetry-derived artifacts (e.g., commands, log fragments), to likely improve situational awareness and support timely decision-making. 
\end{tcolorbox}
}

The \textbf{MITRE ATT\&CK data sources}\footnote{\url{https://attack.mitre.org/datasources/}} describe telemetry collected by sensors (e.g., process creation, file access, authentication logs). Most \textbf{analyst queries focused on interpreting artifacts that these sensors collect, such as commands, scripts, OS processes or file paths}. While subject themes do not directly align with MITRE ATT\&CK data sources, they reflect reasoning shaped by telemetry. For example, a PowerShell command launching a scheduled task maps to \textit{Command Execution}.

\begin{figure}[!t]
    \centering
    \includegraphics[width=1\linewidth]{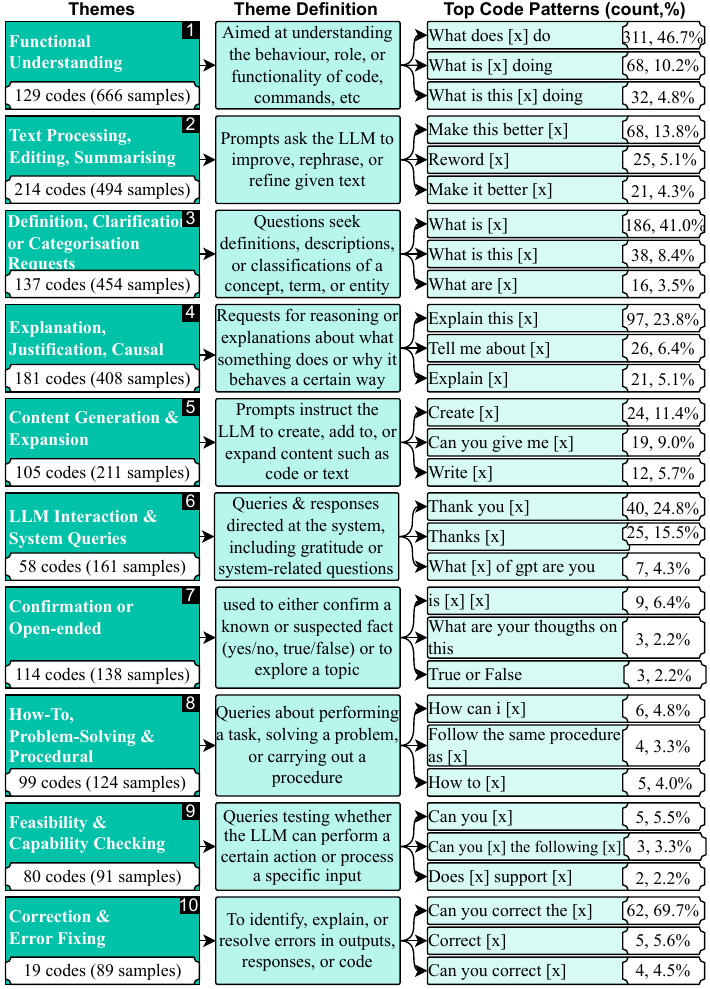}
    \caption{Summary of Pattern Themes (top 10)}
    \label{fig:pattern-themes}
\end{figure}

\subsubsection{SOC Analyst Query Patterns}

Figure~\ref{fig:pattern-themes} shows how analysts linguistically framed their prompts. The most frequent pattern, \textbf{Functional Understanding} (21\%), included queries like \textit{``what does [x] do''}. These were used to interpret unfamiliar commands, scripts, or processes. \textbf{Text Processing and Editing} (16\%) reflected efforts to refine reports or alerts (\textit{``reword this"}), while \textbf{Definition} and \textbf{Explanation} requests (15\% and 14\%) were used to close knowledge gaps. Less common but operationally meaningful patterns included \textbf{Content Generation} (e.g., writing rules or code) and \textbf{Validation/Correction} (e.g., \textit{``is this regex valid?"}). Importantly, these patterns reflect not just style but \textbf{intent}. For example, ``what does'' and ``explain'' queries almost always mapped to command interpretation tasks, while ``make this better'' appeared almost exclusively in text editing. 

{
\begin{tcolorbox}
[width=1\linewidth, center,  left=10pt, right=10pt, top=2pt, bottom=2pt,label=res4,colback=Orchid!10,colframe=Orchid!10,boxrule=0.5pt]
   \textbf{Query phrasing acts as a proxy for analyst reasoning goals, reflecting their underlying intentions and priorities, as well as their problem-solving approach and intended outcomes.}. 
\end{tcolorbox}
}

\subsubsection{Alignment with Professional Cybersecurity Frameworks} 
Building on the task and query subject, we explored whether analyst queries aligned with established cybersecurity roles and competencies. We reviewed all queries against the NICE Framework for Cybersecurity\footnote{\url{https://www.nist.gov/itl/applied-cybersecurity/nice/nice-framework-resource-center}} by matching each task and subject to the NICE categories. This comparison revealed that 93\% of analyst queries could be reasonably associated with at least one NICE Task, Knowledge, or Skill element; the other 7\% mostly included task themes like \textit{LLM Chat} or \textit{Other}. For instance, command interpretation queries align with K0805 (Knowledge of command-line tools and techniques), while email crafting aligns with S0610 (effective communication skills).

{
\begin{tcolorbox}
[width=1\linewidth, center,  left=10pt, right=10pt, top=2pt, bottom=2pt,label=res5,colback=Orchid!10,colframe=Orchid!10,boxrule=0.5pt]
   These findings suggest that \textbf{LLM use was far from arbitrary}; it \textbf{clusters around} competencies already codified by \textbf{industry standards, such as the NICE framework}, suggesting that LLMs could be tailored to support NICE-specific work roles rather than providing generic assistance. 
\end{tcolorbox}
}

\subsection{Conversation Level Analysis}  
\label{sec:results-conversation-phaseV}
One-off interactions comprised 41\% of all interactions; the remaining 59\% formed 532 multi-step conversations. The task themes ordering was stable across interaction types: \textit{Command Understanding/Analysis} was most common, followed by \textit{Text Processing}, then \textit{Code/Script} tasks. 

While the previous section focused on individual queries, this phase examines complete analyst-LLM \textit{conversations}, each defined as a sequence of (temporally) linked queries. This broader lens allows us to examine how analysts pursue and refine their goals over multiple interaction steps. Our analysis examines (1) the distribution and structure of analyst-LLM conversations, highlighting common short and iterative interaction patterns; (2) the ways analysts structure queries and refine their outputs within multi-step sessions; and (3) the task transitions and sequences that characterise real-world analyst workflows when engaging with LLMs. This descriptive lens provides insight into how analysts employ LLMs as flexible, context-sensitive tools for reasoning, drafting, and refinement in SOC operations.

Our thematic coding centred on analyst queries, aligning with our objective of understanding how SOC analysts engage with LLMs in practice. Although we reviewed LLM responses to contextualise conversation-level analyst intent, we did not formally code them. This decision was based on the following considerations: (1) LLM outputs can vary in content or structure even when prompted identically, depending on deployment settings, making systematic analysis less reliable; (2) our primary research objective was to analyse analyst behaviour rather than audit or benchmark the LLM itself; 3) defining response utility is highly analyst- and context-dependent and often lacks objective ground truth; and (4) LLM responses were de-identified, making it difficult to judge the full task context. This analytic focus allowed us to prioritise the human perspective, capturing analyst reasoning patterns and interaction strategies, without over-attributing meaning to the LLM's behaviour, which may vary depending on deployment or model settings.

\subsubsection{Conversation Lengths} 

Table~\ref{tab:conversations} summarises conversation length and the associated mean and median durations (elapsed time between first and last analyst message; LLM response time is not measured). Most conversations were brief: 57\% involved just two analyst steps, and 75\% contained only 2–3 queries. Only a small fraction (just over 4\%) exceeded 10 steps. Durations increase with length, but longer conversations show much higher variance, with mean durations often far exceeding the median, suggesting that many multi-turn sessions are prolonged by idle periods or interruptions rather than sustained engagement. The following sections unpack how tasks and reasoning patterns unfold across these conversation lengths.

\begin{table}[htbp]
\centering
\footnotesize
\caption{Summary of Conversation Lengths and Durations}
\label{tab:conversations}
\begin{tabular}{@{}p{0.8cm}p{2cm}p{0.9cm}p{1.1cm}p{2cm}@{}}
\toprule
Length & \# Conversations & \% of Total & \# Analysts & Mean / Median Duration (minutes) \\
\midrule
2 & 303 & 57.0\% & 35 & 15.8 / 2 \\
3 & 96 & 18.0\% & 22 & 10.2 / 5 \\
4 & 39 & 7.3\% & 18 & 28.9 / 13 \\
5 & 27 & 5.1\% & 10 & 33.1 / 13 \\
6 & 22 & 4.1\% & 13 & 56.4 / 22 \\
7 & 13 & 2.4\% & 8 & 143.4 / 27.5 \\
8 & 8 & 1.5\% & 5 & 145.0 / 89.5 \\
9 & 7 & 1.3\% & 4 & 99.0 / 45.5 \\
10 & 6 & 1.1\% & 5 & 806.2 / 146 \\
11 & 3 & 0.6\% & 3 & 66.0 / 53 \\
12 & 1 & 0.2\% & 1 & 21.0 / 21 \\
13 & 1 & 0.2\% & 1 & 31.0 / 31 \\
14 & 1 & 0.2\% & 1 & 1151.0 / 1151 \\
19 & 1 & 0.2\% & 1 & 49.0 / 49 \\
21 & 1 & 0.2\% & 1 & 298.0 / 298 \\
33 & 1 & 0.2\% & 1 & 5770.0 / 5770 \\
36 & 1 & 0.2\% & 1 & 211.0 / 211 \\
37 & 1 & 0.2\% & 1 & 17728.0 / 17728 \\
\bottomrule
\end{tabular}
\end{table}

\subsubsection{Most Common Two-Step Conversation Sequences} Analysis of frequent two-step sequences shows distinct patterns, with the most common being consecutive \textit{Command Understanding/Analysis} (16.2\% of conversations). This suggests that analysts often explore multiple related commands in succession or seek clarification on different aspects of the same command. Analysts often asked variations of \textit{``What does [x] do?"} in both steps. Another common sequence is \textit{Command Understanding/Analysis} followed by \textit{Summarising Command} (3\%), reflecting a workflow of interpreting and then summarising commands for reporting. 

Several editing-related loops include repeated \textit{Editing/Rewriting Text} (11\%) and iterative revisions for tone or clarity. The third most common loop (4.7\%) is repeated \textit{Code, Script \& Regex Analysis}. Other two-step patterns (1--3\%) involve \textit{Attack \& Threat Intel}, \textit{Tool/Software Understanding}, \textit{Query/Search Related}, and \textit{Email Analysis}.

The recurrence of short, structured sequences suggests that analysts are not simply issuing isolated queries. Instead, they use the LLM to fill in gaps as they piece together the context of an event. For example, repeated command understanding shows how analysts incrementally contextualise related commands, while rewriting or summarising sequences often reflect the need to refine language or tone for reporting or documentation. \textbf{This behaviour reinforces the LLM's role as a flexible tool: analysts use it to interpret telemetry or polish content in small, focused iterations that support the analysts' larger investigative or communicative goals}.

\subsubsection{Three-Step Conversation Sequences}

In three-step conversations, analysts often stayed on one theme, e.g., \textit{[Editing/Rewriting Text, Editing/Rewriting Text, Editing/Rewriting Text]} (5\%), making iterative refinements like \textit{``Make this better''}, then \textit{``Make it shorter''}, and \textit{``Make it sound better''}, showing that the LLM was used as a drafting and refinement tool for alerts, notes, or emails. Triplets of command understanding queries (4\%) often repeated prompts like \textit{``What does [x] do?''} or \textit{``Explain [x]''}, showing iterative probing to interpret commands or components.

A small number of sequences involved transitions between different task themes (e.g., from command analysis to text editing, or alternating between command and tool understanding), highlighting cases where the analyst first interprets a command and then translates that understanding into an incident narrative. \textbf{These three-step interactions reveal how analysts engage the LLM to incrementally make sense of complex information, refine their articulation, and support analytical workflows}. The sequences reflect real-world cognitive processes, particularly when dealing with uncertainty or unfamiliar commands or scenarios.

\begin{figure*}[ht]
    \centering
    \includegraphics[width=0.9\linewidth]{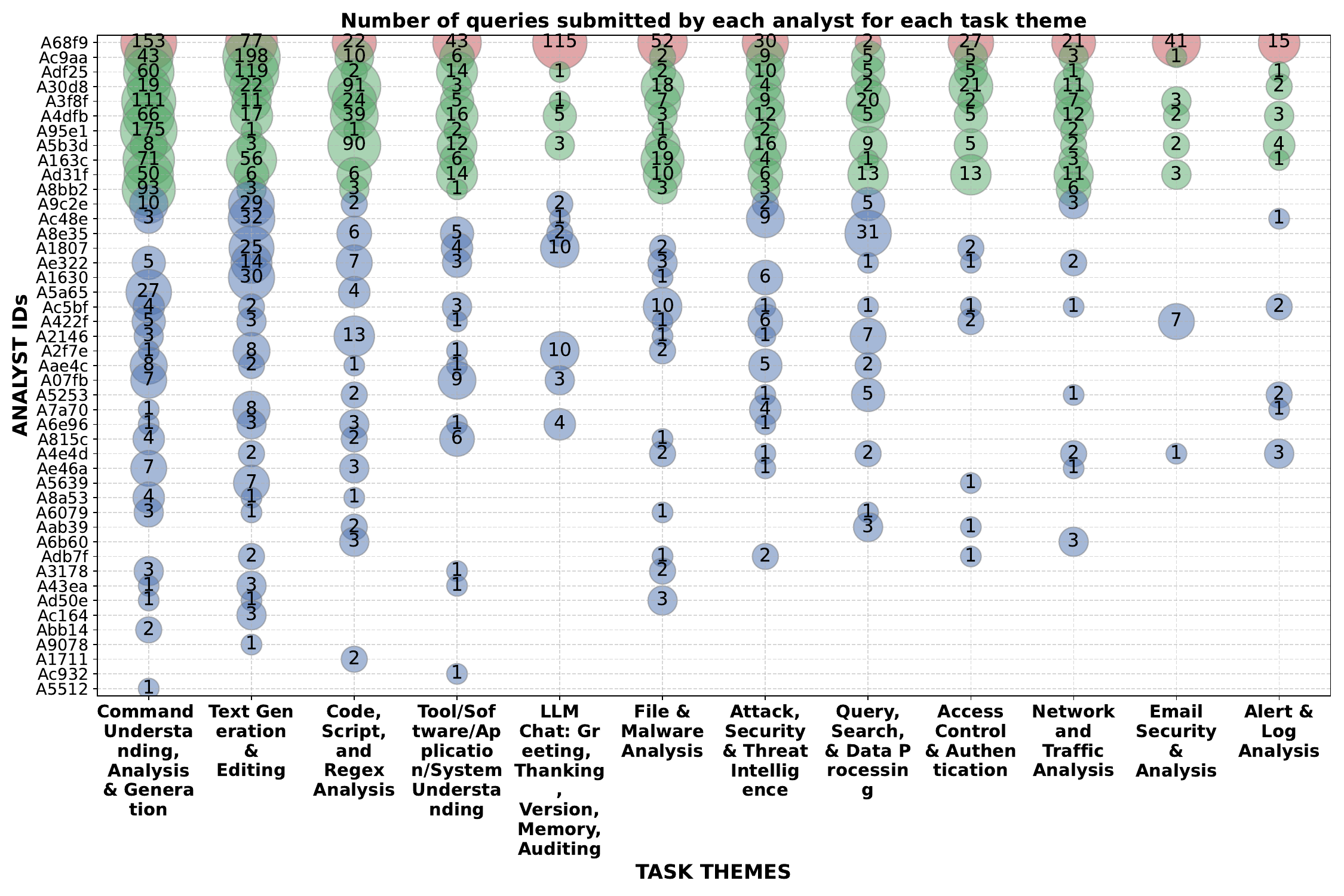}
    \caption{Number of queries per analyst (ordered by activity) and task theme (ordered by frequency). Analyst clusters are colour-coded: red for most active, green for moderate, and blue for low-usage analysts (same as Figure~\ref{fig:llm_analyst_engagement_summary})}
    \label{fig:analyst_theme_balloon_plot}
\end{figure*}

\subsubsection{Longer Conversations} Most interactions were short (2–3 turns), but longer ones, though rarer, show how analysts build on prior queries to deepen insight and refine results.

These longer conversations spanned a wide range of tasks. Some involved exploratory learning or upskilling, such as understanding machine learning fundamentals or understanding authentication mechanisms. Others reflected hands-on problem-solving, such as writing or debugging scripts, analysing remote shell behaviour, analysing registry modifications, or investigating potentially malicious behaviour (e.g., detecting potential ransomware). A distinct subset of longer conversations centred on professional development, likely related to CVs or performance reviews. These cases highlight the LLM's use beyond core SOC functions, supporting analysts in broader professional contexts. 

Conversations with four or more consecutive \textit{Command Understanding/Analysis} steps suggest thorough exploration and careful interpretation of related commands in high-stakes environments. Other sequences chain distinct but related subtasks, combinations such as \textit{File Understanding} $\rightarrow$ \textit{Command Understanding/Analysis}  $\rightarrow$ \textit{Threat Analysis}  $\rightarrow$ \textit{Summarise LLM Output} indicate layered workflows: first clarifying data or command behaviour, then assessing threats, and finally seeking help with incident reporting. We also observed iterative development conversations with repeated cycles of \textit{Code Writing and Understanding}, \textit{Understanding Errors}, and \textit{Query Writing}, suggesting that LLM supports implementation, error checking, refinement, and formatting. Several multi-step interactions focused on communication tasks—\textit{editing text}, \textit{improving clarity}, and \textit{crafting empathetic emails}, combining structural edits with tone adjustments to produce polished, context-appropriate messages for diverse stakeholders.

\vspace*{-2\baselineskip}

{
\begin{tcolorbox}
[width=1\linewidth, center,  left=10pt, right=10pt, top=2pt, bottom=2pt,label=res6,colback=Orchid!10,colframe=Orchid!10,boxrule=0.5pt]
   Across these conversations, \textbf{patterns suggest that SOC analysts often use LLMs as iterative, context-sensitive tools for reasoning, drafting, and refinement}. Most conversations are short and focused, typically 2--3 steps, centred on tasks such as command understanding, text editing or coding tasks. While longer sessions are rare, they reveal more complex workflows, indicating that LLMs can support layered analytical and communicative goals. These findings suggest emerging usage patterns, though they should be interpreted with caution, given the predominance of brief exchanges and the absence of direct cognitive measures capturing analysts' reasoning processes.
\end{tcolorbox}
}

\subsection{Analyst Level Analysis}

Engagement with the LLM varied widely across analysts in query volume and thematic breadth. Analysts clustered into three groups: \textit{low-usage} (Cluster 0, $n=34$), \textit{moderate} (Cluster 1, $n=10$), and a single \textit{high-volume outlier} (Cluster 2, $n=1$).

We examined daily LLM query activity across all analysts (Figure~\ref{fig:analyst_theme_balloon_plot}) and observed distinct engagement patterns. One analyst (A68f) consistently used the system at a high volume across multiple months. A group of moderate users (e.g., Ac9a, A5b3) engaged in concentrated bursts of activity, likely linked to specific investigations. In contrast, most analysts issued only a few queries on scattered days, indicating sporadic or task-specific use. These behavioural differences support cluster-based segmentation and highlight varied LLM integration styles.

Analysts varied markedly in the task domains they engaged with through the LLM. The most prolific user, \textit{A68f}, demonstrated broad usage across multiple domains, particularly \textit{Command Understanding, Analysis \& Generation}, \textit{LLM Chat/System Queries}, and \textit{Text Generation \& Editing}. This suggests deep LLM integration into technical investigations and communication-related tasks. In contrast, \textit{Ac9a} concentrated overwhelmingly on \textit{Text Generation \& Editing}, with limited engagement in command analysis, indicating a primarily communicative or documentation-focused use.

\textit{A30d} and \textit{A5b3} mainly focused on \textit{Code, Script, and Regex Analysis}, with moderate exploration of other tasks. \textit{Adf2} showed a balanced profile across \textit{Text Generation}, \textit{Command Understanding}, and \textit{Tool/System Understanding}, using the LLM for both exploratory and functional support. \textit{A4df} combined command and code tasks with moderate \textit{Query/Data Processing} and \textit{File Analysis}, indicating technically grounded use. Among lower-volume users, such as \textit{A180}, \textit{Ac48}, and \textit{A95e}, task engagement was more narrowly focused, typically confined to one or two functional categories such as \textit{Text Editing} or \textit{Command Understanding}. Notably, the recurring appearance of \textit{Command Understanding} and \textit{Text Generation} across many analysts underscores their central role in LLM support within SOC workflows.

\begin{figure*}[ht]
    \centering

\includegraphics[width=\linewidth]{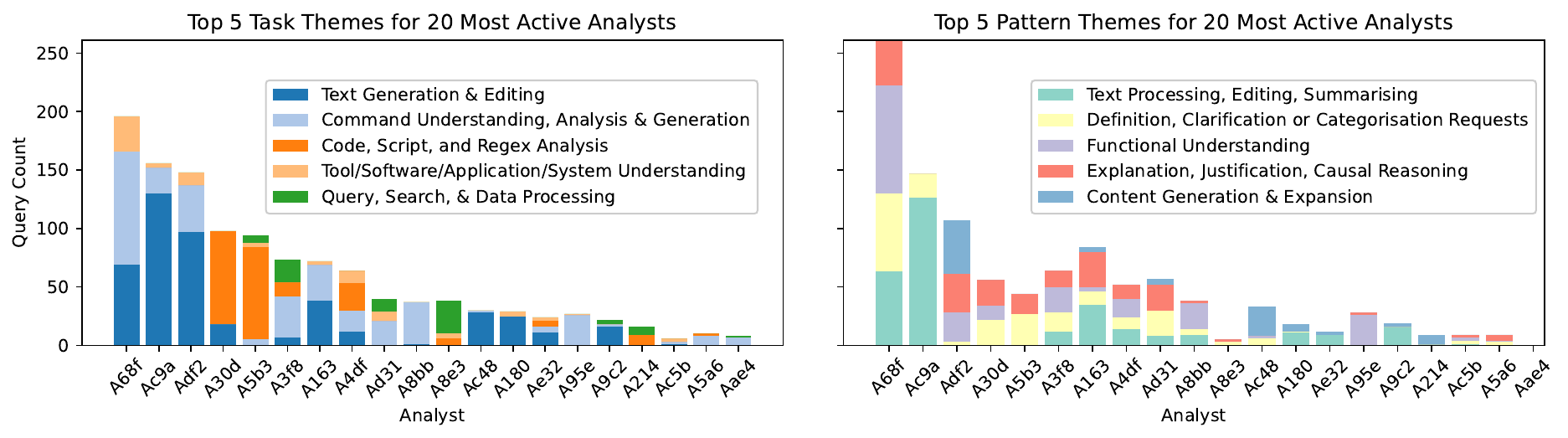}
    \caption{Top 5 task themes (left) and pattern themes (right) used by the 20 most active analysts.}
    \label{fig:side_by_side_task_pattern_legends_sides}
\end{figure*}

These task distinctions are mirrored by varying cognitive strategies at the pattern level. Figure~\ref{fig:side_by_side_task_pattern_legends_sides} shows the top 5 task themes (left) and top 5 pattern themes (right) used by the 20 most active analysts. On the left side, usage is dominated by \textit{Command Understanding} and \textit{Text Generation \& Editing}, though the mix varies by analyst. On the right, \textit{Text Processing, Editing, \& Summarising} is the most prevalent interaction pattern, but analysts also use a mix of patterns.

\textit{A68f}, the most prolific user, led in \textit{Functional Understanding} and \textit{Clarification Requests}, with heavy use of \textit{LLM Interaction \& System Queries}. This pattern aligns with their diverse technical workload and diagnostic dialogue style. In contrast, \textit{Ac9a} showed a documentation-focused usage profile via \textit{Text Processing, Editing \& Summarising}, alongside frequent clarification queries. \textit{Adf2} exhibited an iterative refinement strategy, issuing \textit{Correction \& Error Fixing} and \textit{Content Generation \& Expansion} queries; a cycle of generating, testing, and refining outputs. 

Distinct pattern signatures also appear to reflect role-based tendencies. \textit{A30d} and \textit{A5b3}, for example, consistently used the LLM for \textit{Feasibility Checking}, \textit{How-To Reasoning}, and \textit{Causal Explanation}, aligning with their hands-on technical work and exploratory debugging. Conversely, low-volume users such as \textit{A07f}, \textit{A2f7}, and \textit{Aae4} issued queries primarily related to \textit{Clarification} or \textit{Functional Understanding}, suggesting that for many, the LLM functioned as a reference assistant and a reasoning partner. However, it is important to note that a comprehensive understanding of analysts' specific tasks and context is necessary to accurately interpret such trends; variability in queries may simply reflect the evolving nature of investigative work.

{
\begin{tcolorbox}
[width=1\linewidth, center,  left=5pt, right=5pt, top=1pt, bottom=1pt,label=res7,colback=Orchid!10,colframe=Orchid!10,boxrule=0.5pt]
   We observed that \textbf{LLMs are not used uniformly} across analysts. Instead, \textbf{analysts appear to adapt the tool to their task demands, specific needs, and investigative styles}, ranging from quick clarification and script debugging to collaborative drafting and iterative refinement. The \textbf{LLM successfully supported this diversity without requiring extensive prompt engineering}. This highlights that a well-configured, domain-aligned LLM can support a broad range of task types and prompt styles, enabling quick and effective interaction, even without deep prompting expertise. 
\end{tcolorbox}
}

\subsubsection{Discussion on Potential Disengagement Patterns}
The following analysis examines behavioural patterns among SOC analysts with lower engagement levels to understand where, why and how disengagements occur. We observed two distinct groups of low-engagement analysts.

\textbf{Group 1: Low-Engagement, Early Drop-off Analysts (n = 13)}
These analysts interacted for only 1-2 months. Most (9 out of 13) showed no signs of frustration, implying exploratory or episodic use rather than dissatisfaction. However, three analysts exhibited potential breakdowns that coincided with their last few queries.

\textit{A2f7} repeatedly queried the model version (perhaps unmet transparency need), \textit{A5a6} received two consecutive erroneous answers, with a misclassification ending the session, and \textit{A6e9} disengaged immediately after the LLM refused a request. Another analyst (\textit{Ae46}) ceased engagement after several command analysis queries, although no clear failure pattern was identified. Further analysis revealed that task theme rankings for this group mirrored the broader pool of analysts, indicating that early drop-off was not task-specific. Collectively, this group suggests that \textbf{early disengagement occurred for multiple reasons: in some cases without clear frustration, and in others following specific model failures}.

\textbf{Group 2: Persistent but Mixed-Experience Analysts (n = 6)}
These analysts remained active for four to eight months. Two analysts (\textit{Ae32}, \textit{Aae4}) experienced apparent errors yet persisted, indicating that perceived utility can offset isolated failures. Other patterns suggest diverse usage motives: \textit{A9c2} moved past an early debugging issue; \textit{A214} relied on the LLM for repeated Sumo Logic refinements; \textit{A8bb} may have engaged only when necessary; and \textit{A525} mixed non-SOC queries with SOC ones. \textbf{This group suggests that analysts tolerate occasional mistakes, perhaps if using the LLM for specific tasks is beneficial.} While these patterns are informative, future work should observe SOC analysts and conduct interviews to validate disengagement reasons.

\subsubsection{AI-Assisted Decision-Making Perspective}

Only a small proportion of analysts explicitly sought \textit{recommendations} (we use recommendations here to mean classification requests, such as, \textit{``is it malicious"} or \textit{``is this a threat"}) from the LLM. Across the dataset, approximately 4\% of all queries involved requests for such binary judgements.
For example, if we only consider \textit{Command Understanding, Analysis \& Generation}, in total there were 950 queries (31\%), and only 3\% involved explicit recommendation requests, and these were issued by just seven analysts. 

{
\begin{tcolorbox}
[width=1\linewidth, center,  left=5pt, right=5pt, top=1pt, bottom=1pt,label=res8,colback=Orchid!10,colframe=Orchid!10,boxrule=0.5pt]
   This suggests that \textbf{\textit{most} analysts do not seek recommendations} from the LLM. Instead, they appear to \textbf{prefer evidence} or \textbf{context} that allows them to understand what is happening and \textbf{\textit{make decisions} independently}. 
\end{tcolorbox}
}

To effectively support SOC analysts, AI systems must align with this preference for evidentiary support over prescriptive output, for example, by using \textit{machine-in-the-loop} approach~\cite{Miller2023-uz}, which presents evidence for and against outcomes rather than definitive answers to support informed decisions and preserve human agency~\cite{Le2024-mc, Shneiderman2022-uj}. However, because some analysts prefer explicit recommendations, a one-size-fits-all approach risks leaving needs unmet. Therefore, investigating the need for adaptive systems that support both interpretive and directive workflows is essential.

\section{Discussion}
\label{sec:discussion}

This study set out to investigate a core question (restated briefly): \textit{How do SOC analysts utilise LLMs in their daily workflows, including the specific tasks they apply them to, how these uses align with established cybersecurity frameworks (e.g., NICE), and the conversational patterns that characterise their interactions?} Below, we address this question through a multi-level analysis and conclude with a discussion of the study's limitations and the broader implications for designing LLM-based assistance in SOCs.

\subsection{Query Level: Analyst Needs and Task Focus}
\label{sec:discussion-rq1}

The progressive rise in LLM use across tasks (Figure~\ref{fig:analyst_theme_balloon_plot}) and sustained engagement by more analysts (Figure~\ref{fig:llm_analyst_engagement_summary}) show that \textbf{analysts gradually integrated the LLM into daily workflows}, applying it to more tasks over time (see Figure~\ref{fig:month_theme_balloon_plot} in Appendix). These usage patterns suggest that for some analysts, the \textbf{LLM delivered sufficient value to merit repeated use during investigative sessions}.

\textbf{The LLM fulfilled a key, hidden interpretive layer missing from current SOC tooling, particularly in interpreting low-level telemetry}, such as commands, files, system processes and scripts. This highlights a key function of the LLM: acting as a real-time interpreter or explainer of low-level telemetry. During investigations, Tier-1 analysts need to make sense of raw telemetry~\cite{Kersten2023-wy}. Instead of relying solely on documentation or prior knowledge, analysts used LLMs as on-demand aids to translate such artifacts into actionable context, potentially helping analysts discern whether observed activity was benign or malicious \cite{Kersten2023-wy}. In this way, LLMs appear to reduce time-consuming lookups, ease cognitive load, and improve situational awareness. 

\textbf{Text processing use cases suggest a role in enhancing operational efficiency}. Rather than spending time on repetitive communication or documentation tasks, analysts utilised the LLM to generate more polished and comprehensive materials for reporting and sharing. This would have offloaded non-investigative cognitive effort, supporting productivity without disrupting core investigative flow.

\textbf{LLM usage is overwhelmingly concentrated in command interpretation and text processing}, with most other task types showing signs of increasing adoption, but there is only preliminary evidence in our dataset to claim widespread use. Command and text processing tasks are used continuously and with growing frequency across analysts and months (see Figure \ref{fig:month_theme_balloon_plot} in the Appendix).

\textbf{Query patterns reflected clear cognitive intent}. Analysts' queries were not just syntactic variations but reflected clear cognitive intent. Functional queries like \textit{``what does [x] do''} mapped to interpretive tasks such as command understanding, while requests like \textit{``reword this''} reflected text refinement goals. This alignment between linguistic patterns and task types suggests that query structure can serve as a proxy for the analyst's reasoning objective.

{
\begin{tcolorbox}
[width=1\linewidth, center,  left=2pt, right=2pt, top=1pt, bottom=1pt,label=design3,colback=Orange!10,colframe=Orange!10,boxrule=0.5pt]

\textbf{Design Implication:} Surface-level telemetry is not the same as true comprehension. Embedding plain-language, \emph{just-in-time} explanations into analyst workflows could close this gap. For example, \textbf{embedding analyst role- and history-adaptive LLM explanations in SOC dashboards could accelerate triage, reduce cognitive load, and support evidence-based decisions}. Prior research highlights the need for actionable, context-rich explanations for SOC analysts~\cite{Mink2023-ea}, and that user-tailored interfaces improve decision accuracy and satisfaction~\cite{Lai2023-hj}. 
Agentic RAG~\cite{Xu2025-pc} could be explored to enable such explanations by dynamically tailoring retrieval and generation based on the analyst's role and recent interactions, ensuring that each explanation is contextually relevant and personalised to the analyst's expertise and workflow.

\end{tcolorbox}
}

\subsection{Conversation-level: Reasoning Sequences and LLM Roles}

\textbf{Our analysis reveals that SOC analysts engage with LLMs through brief, focused, and task-driven conversations}. The results indicate that these 2-3 step conversations are not exploratory; rather, they are concise, goal-directed exchanges grounded in real-time investigative needs. Such behaviour aligns with the notion of \textit{assisted intelligence}~\cite{Jiang2024-eq}, in which LLMs are used as on-demand tools for clarification, explanation, or transformation.

\textbf{In operational SOCs, LLMs act as lightweight, on-demand aides rather than persistent copilots}, helping analysts reach objectives without elaborate prompts. This real-world perspective contrasts with prior work focused on technical capabilities or prototypes~\cite{Freitas2024-sx, Siracusano2023-qs, Hassanin2025-eo}, highlighting the need to study LLMs in live environments.

\textbf{LLM use in our dataset aligns closely with established cognitive and collaborative AI role frameworks}. For instance, command understanding aligns with the \textit{clarifier}~\cite{Eigner2024-zd}, the \textit{Assisted Intelligence} role~\cite{Jiang2024-eq}, and the \textit{consultant/search engine} archetypes~\cite{Song2024-mw}. Text processing tasks correspond to the design phase in~\cite{Eigner2024-zd}, and the \textit{Augmented Intelligence} category~\cite{Jiang2024-eq}. Meanwhile, iterative reasoning and hypothesis testing interactions resemble the \textit{deliberative partner} model~\cite{Ma2024-uh}, \textit{ExtendAI} role~\cite{Reicherts2025-we} and align with \textit{Cooperative Intelligence}~\cite{Jiang2024-eq}. Together, these roles underscore the varied, situational ways LLMs support cognition within the SOC environment.

{
\begin{tcolorbox}
[width=1\linewidth, center,  left=2pt, right=2pt, top=1pt, bottom=1pt,label=design4,colback=Orange!10,colframe=Orange!10,boxrule=0.5pt]

\textbf{Design Implication:} Embed LLMs directly into SOC dashboards \textbf{so analysts can summon help for specific microtasks} (e.g., interpreting a suspicious command, summarising related logs, or drafting an incident note) without leaving their current workflow to \textbf{reduce context switching}, and suggest logical next steps (for example, offering a summary after several explanation requests).

\end{tcolorbox}
}

{
\begin{tcolorbox}
[width=1\linewidth, center,  left=2pt, right=2pt, top=1pt, bottom=1pt,label=research1,colback=LimeGreen!10,colframe=LimeGreen!10,boxrule=0.5pt]

\textbf{Research Opportunity:} While most analyst-LLM interactions focused on single tasks, some involved task transitions (e.g., command analysis to summarisation), raising questions about the value of task-specific LLMs and how to support seamless switching. Future studies could assess if task-specific LLMs enhance speed, transparency, and trust in SOC decision-making. Furthermore, \textit{Agentic AI} could be explored to reduce context switching by assigning specialised agents to key micro-tasks within the analyst's workflow~\cite{Sapkota2025-ry}, and even dynamically incorporate expert analysts~\cite {Xu2025-pc}, while carefully managing the security risks and operational complexity~\cite{Kshetri2025-gq}.
   
\end{tcolorbox}
}

\subsection{Analyst-Level: Usage Patterns and Integration Approaches}

Our analyst-level analysis reveals key differences in LLM usage styles, cognitive strategies, and decision preferences, offering critical implications for the design, deployment, and evaluation of AI-powered SOC tooling.

\textbf{LLM use varies.} Analysts differed not only in the frequency of LLM engagement but also in the types of tasks they delegated and how they framed their requests. Some used LLMs for command parsing and debugging; others focused on editing or summarising. These distinctions suggest an opportunity for investigating the effectiveness of role-aware, adaptive integration strategies that personalise tooling based on emerging usage patterns. Our findings build on early claims of LLMs as cognitive aides and validate emerging theories of diverse user strategies and interaction modes~\cite{Freitas2024-sx, Siracusano2023-qs,guo2024investigating, Pang2025-vk}. While prior work explored LLMs for structured extraction and task acceleration, we show that usage patterns naturally align with analysts' task orientations, without prompt engineering or adaptation.

\textbf{LLMs function primarily as evidence interpreters, not decision-makers.} Only 4\% of queries sought explicit recommendations (e.g., ``is this malicious?"). Instead, most analysts requested explanation, interpretation, or contextualisation, highlighting a strong preference for maintaining decision authority. Our findings support designing LLMs as evaluative aides that surface evidence without prescribing actions, aligning with concerns about over-reliance on automation and reinforcing the need for expert-centred, machine-in-the-loop systems~\cite{Bucinca2021-di,Miller2023-uz, Le2024-mc, Shneiderman2022-uj}. Analysts’ preference for interpretation over recommendations warrants further study.

\subsection{Other factors that modulate routine engagement: isolated errors and trust.} 

\vspace*{-0.4\baselineskip}

\textit{Disengagement is often episodic, not systemic.} 
Among low-usage analysts, disengagement typically followed isolated errors or unmet expectations (e.g., refusal). Others continued using the LLM despite early failures, indicating that value outweighs occasional breakdowns. Clarifying outputs or allowing retries may reduce drop-off.

\textit{Sustained engagement hints at emerging trust}. Trust underpins effective human-AI teaming, shaping user engagement over time. Prior work distinguishes \textit{situational trust} (based on immediate interactions) from \textit{learned trust} (built from experience)~\cite{Duan2024-ih,McGrath2024-dr,Marsh2005-gp}. In our dataset, repeated use of the LLM for complex interpretive tasks, especially by some analysts, suggests the latter, where users expect reliable support in specific contexts. However, without direct feedback, we cannot confirm trust as the driving factor. Future research should combine usage data with trust metrics to clarify the nature of trust in human-AI collaboration.

{
\begin{tcolorbox}
[width=1\linewidth, center,  left=2pt, right=2pt, top=1pt, bottom=1pt,label=design1,colback=Orange!10,colframe=Orange!10,boxrule=0.5pt]

\textbf{Design Implication:} \textbf{Surface \emph{evidence}, not \textit{recommendations} for investigative tasks}~\cite{Miller2023-uz,Le2024-mc,Reicherts2025-we}.  Because only 4\% of analyst queries requested explicit malicious/benign judgements, for tasks that request such information, interfaces should default to an \textit{evidence-first} mode, highlighting log excerpts, command traces, and ATT\&CK linkages while omitting prescriptive labels.  
   
\end{tcolorbox}
}

{
\begin{tcolorbox}
[width=1\linewidth, center,  left=2pt, right=2pt, top=1pt, bottom=1pt,label=design2,colback=LimeGreen!10,colframe=LimeGreen!10,boxrule=0.5pt]

\textbf{Research Opportunity:} Future studies could investigate whether the preference for evidence-oriented outputs (4\% in our study) generalises to different settings, and also compare the impact of evidence-oriented versus recommendation-oriented outputs on accuracy, response time, and calibrated trust~\cite{Fogliato2022-dj,Bucinca2021-di} to determine which support style minimises misuse yet maintains efficiency during high-pressure triage.
   
\end{tcolorbox}
}

\vspace{2pt}

{
\begin{tcolorbox}
[width=1\linewidth, center,  left=5pt, right=5pt, top=1pt, bottom=1pt,label=resarchover,colback=Cerulean!15,colframe=Cerulean!15,boxrule=0.5pt]
   \textbf{Answering the RQ:} SOC analysts primarily use LLMs as cognitive aids to interpret technical artifacts and streamline communication using brief, goal-oriented queries that are closely aligned with the NICE Workforce Framework and MITRE ATT\&CK Framework. Most analysts view the LLM as a cognitive aid rather than a decision-maker. \textbf{Our findings indicate that LLMs serve as flexible, on-demand aids that enhance rather than replace analyst expertise}. Analysts appear to optimise both time and LLM utility through short, high-value interactions, quickly interpreting obfuscated telemetry for situational awareness~\cite{Kersten2023-wy} and efficiently completing writing tasks, while retaining full decision-making authority.
\end{tcolorbox}
}

\subsection{Limitations and Future Work}

This study has several limitations. First, all data were collected from a single SOC using a specific LLM deployment. SOCs vary in tools, workflows, and policies, so our findings may not generalise to other operational contexts or different LLMs. Second, we lack objective performance metrics, such as triage speed, false‐positive rates, LLM response accuracy, or overall investigation accuracy. However, it is important to note that the primary goal of this study was not to evaluate performance outcomes but to explore how analysts interacted with the LLM in practice. High engagement levels suggest analysts found the LLM useful, but we cannot conclude whether these interactions translated into better or faster decisions. Third, there may be a novelty effect, with some analysts engaging with the LLM simply because it was new or conveniently accessible, rather than due to demonstrated utility, and some may have avoided it because of the perceived hallucination issues. As such, it is hard to separate genuine productivity gains from curiosity-driven use without further investigation. 

Future work can build on our findings in several ways. First, replicating this study across multiple SOCs with different toolchains, team sizes, and operating procedures will help assess the generalisability of observed interaction patterns and support the use of objective measures, such as time‐to‐triage. Second, integrating LLM features directly into SOC dashboards could reduce context-switching and offer deeper insight into how interface design shapes analyst-LLM interaction. Finally, longitudinal studies that track new and experienced analysts over time can reveal how trust, adoption, and productivity evolve as the novelty of the LLM wears off. Future longitudinal studies should include interviews to capture analysts' experiences and perspectives, integrating these insights into design guidelines for effective LLM-analyst collaboration. By pursuing these directions, future work can move from descriptive usage patterns and provide more comprehensive insights into the role of LLMs in supporting analyst decision-making and collaboration within operational SOCs.

\vspace{5pt}
\noindent\textbf{Ethical Considerations:} Ethical approval for this study was granted by the researchers' host organisation's Ethics and Privacy Office (Approval \#: 025/25). Analysts were informed that their interactions would be anonymously logged to understand how they engaged with the new LLM tools, and consented to this. Informed consent was obtained by the industry partner, and participation was completely voluntary. Prior to data sharing, the industry partner de-identified the data and removed sensitive information. All procedures were reviewed and approved by both the industry partner and the host organisation's ethics and privacy teams.

\section{Conclusion}
\label{sec:conclusion}

This paper explored how SOC analysts interact with LLMs in an operational setting. By analysing thousands of analyst-generated queries, we found that analysts use LLMs as on-demand, task-focused cognitive aids for a variety of tasks, including explaining commands, writing scripts, or improving documentation, rather than as full-time copilots. Future tools could embed LLMs directly into SOC systems to deliver adaptive, context-sensitive assistance. From a research perspective, our work provides a grounded foundation for understanding analyst-LLM collaboration and underscores the need for outcome-based evaluations of their real-world impact on SOC decision-making. 

\section{Acknowledgments}

This work is supported by the Commonwealth Scientific and Industrial Research Organisation (CSIRO) Collaborative Intelligence (CINTEL) Future Science Platform (FSP) and eSentire, Inc. Figures are designed using icons from Flaticon.com.

\bibliographystyle{IEEEtran}
\bibliography{references}

\appendix

\section{Paper Appendix: Themes and Additional LLM Usage Patterns}

\begin{figure}[h!]
    \centering
    \includegraphics[width=0.9\linewidth]{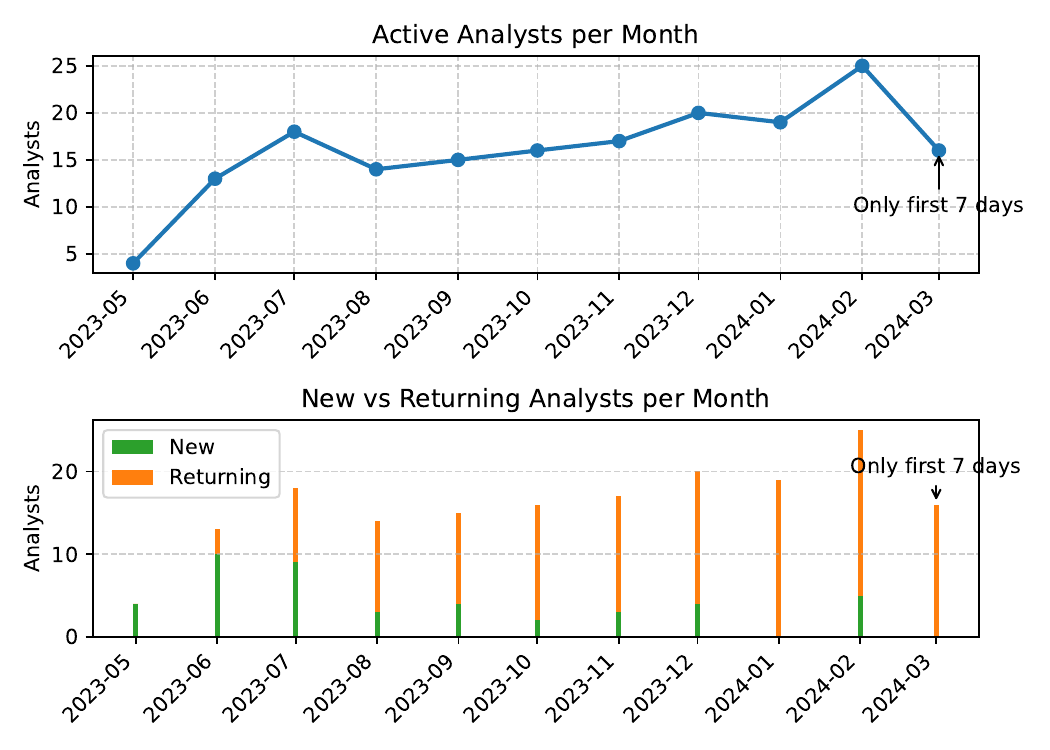}
    \caption{Analysts' engagement over months.}
    \label{fig:analyst_engagement_metrics}
\end{figure}

\textbf{LLM Usage Summary: Monthly Patterns. } Figure~\ref{fig:analyst_engagement_metrics} summarises SOC analyst engagement with the LLM system, showing monthly active users (top) and new vs. returning users (bottom), highlighting adoption trends and increasing integration into daily workflows.

\begin{figure*}[ht]
    \centering
    \includegraphics[width=0.85\linewidth]{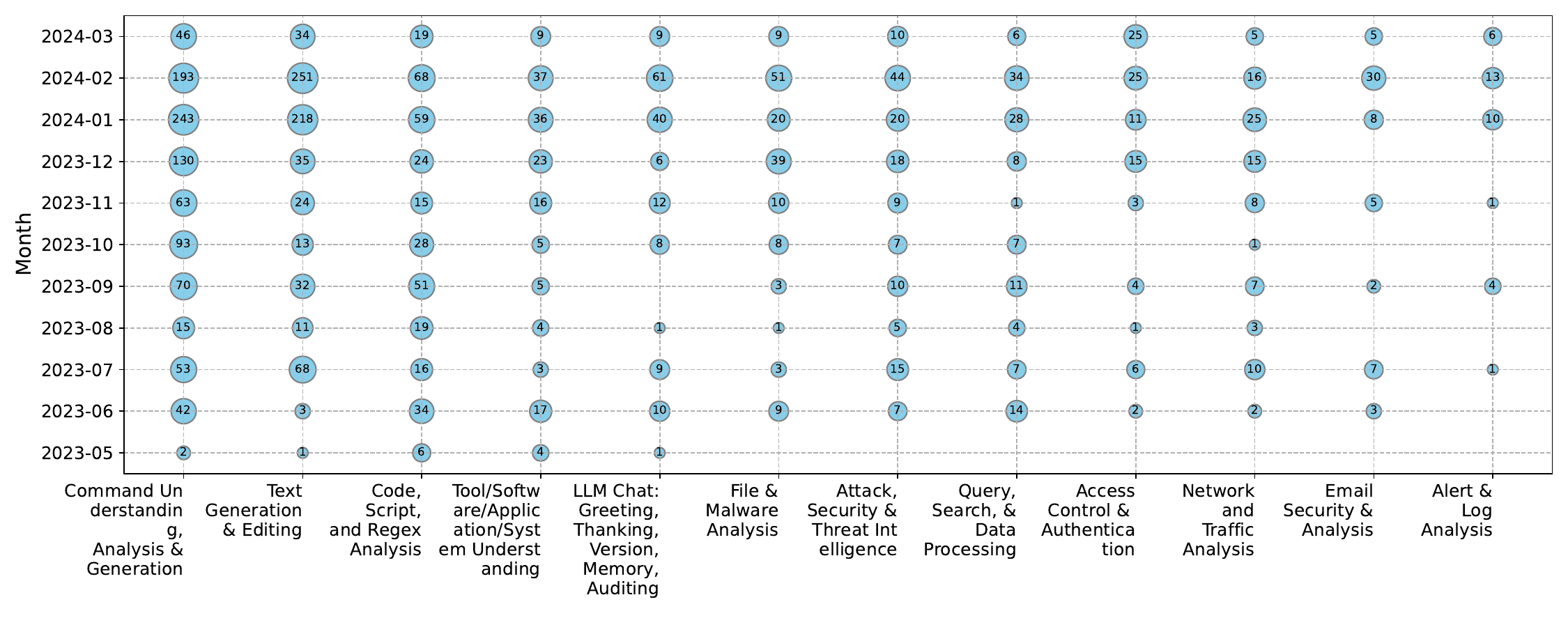}
    \caption{Monthly query volume by task theme.}
    \label{fig:month_theme_balloon_plot}
\end{figure*}

\textbf{Monthly Task‐Theme Usage. } Figure~\ref{fig:month_theme_balloon_plot} shows monthly query volumes by Task Theme, highlighting broad growth that suggests LLMs have become multi-role assistants in SOC investigations, reporting, and automation.

\begin{table*}[htbp]
\centering
\caption{Summary of Pattern Theme with Top 3 Codes}
\begin{tabular}{|p{2.5cm}|p{7.0cm}|p{5.5cm}|p{1.0cm}|}
\hline
\textbf{Theme} & \textbf{Theme Definition} & \textbf{Top 3 Code Patterns (Count, \%)} & \textbf{\# Codes (Total)} \\ \hline
Functional Understanding & Queries aimed at understanding the behaviour, role, or functionality of code, commands, etc & what does [x] do (311, 46.7\%), what is [x] doing (68, 10.2\%), what is this [x] doing (32, 4.8\%) & 129 (666) \\ \hline

Text Processing, Editing, Summarising & Prompts ask the LLM to improve, rephrase, or refine given text & make this better [x] (68, 13.8\%), reword [x] (25, 5.1\%), make it better [x] (21, 4.3\%) & 214 (494) \\ \hline

Definition, Clarification or Categorisation Requests & Questions seek definitions, descriptions, or classifications of a concept, term, or entity & what is [x] (186, 41.0\%), what is this [x] (38, 8.4\%), what are [x] (16, 3.5\%) & 137 (454) \\ \hline

Explanation, Justification, Causal Reasoning & Requests for reasoning or explanations about what something does or why it behaves a certain way. & explain this [x] (97, 23.8\%), tell me about [x] (26, 6.4\%), explain [x] (21, 5.1\%) & 181 (408) \\ \hline

Content Generation \& Expansion & Prompts instruct the LLM to create, add to, or expand content such as code or text & create [x] (24, 11.4\%), can you give me [x] (19, 9.0\%), write [x] (12, 5.7\%) & 105 (211) \\ \hline

LLM Interaction \& System Queries & Queries and responses directed at the system itself, including gratitude or system-related questions & thank you [x] (40, 24.8\%), thanks [x] (25, 15.5\%), What [x] of gpt are you (7, 4.3\%) & 58 (161) \\ \hline

Confirmation or Open-ended & used to either confirm a known or suspected fact (yes/no, true/false) or to explore a topic & is [x] [x] (9, 6.5\%), what are your thougths on  [x] (3, 2.2\%), [x] True or False (3, 2.2\%) & 114 (138) \\ \hline

How-To, Problem-Solving \& Procedural & Queries about performing a task, solving a problem, or carrying out a procedure & how can i [x] (6, 4.8\%), how to [x] (5, 4.0\%), in [x] pick [x] (4, 3.2\%) & 99 (124) \\ \hline

Feasibility \& Capability Checking & Queries testing whether the LLM can perform a certain action or process a specific input & can you [x] (5, 5.5\%), can you [x] the following [x] (3, 3.3\%), does [x] support [x] (2, 2.2\%) & 80 (91) \\ \hline

Correction \& Error Fixing & To identify, explain, or resolve errors in outputs, responses, or code & can you correct the [x] (62, 69.7\%), correct [x] (5, 5.6\%), can you correct [x] (4, 4.5\%) & 19 (89) \\ \hline

Data Analysis and Interpretation & Prompts focused on decoding, interpreting, or drawing insights from data or patterns & decode [x] (15, 22.7\%), how is [x] (3, 4.5\%), are these [x] same (2, 3.0\%) & 48 (66) \\ \hline

Information Gathering \& Retrieval & Requests for factual information, examples, or references related to a topic or concept & tell me about [x] (2, 3.4\%), Is there [x] (2, 3.4\%), give me [x] (2, 3.4\%) & 52 (58) \\ \hline

Explore Options \& Alternatives & Questions that explore multiple possibilities, compare options, or consider different approaches to a problem, decision, or scenario. Often involves weighing alternatives or identifying the most appropriate course of action. & which of the following [x] (2, 3.8\%), i need [x] for the [x] (1, 1.9\%), which [x] patched [x] (1, 1.9\%) & 52 (53) \\ \hline

Event \& Outcome Analysis & Aimed at interpreting security events, determining what occurred, assessing impact, or understanding post-activity implications (e.g., after a login, installation, or process execution) & what is the outcome of [x] (6, 15.8\%), whats happening here [x] (4, 10.5\%), what is the result of [x] (3, 7.9\%) & 27 (39) \\ \hline

Observation and Analyst Feedback & Statements or follow-up queries where analysts reflect on or react to prior LLM responses, sometimes providing their own assessment, confirming correctness, or probing further clarification & im trying to [x] could [x] help with that (1, 2.6\%), but the previous version of the [x] worked (1, 2.6\%), I can't find any information to verify [x] (1, 2.6\%) & 38 (38) \\ \hline

\end{tabular}
\end{table*}

\begin{table*}[htbp]
\centering
\caption{Summary of Task Theme with Top 3 Codes}
\begin{tabular}{|p{2.9cm}|p{5.65cm}|p{6.0cm}|p{1.0cm}|}
\hline
\textbf{Theme} & \textbf{Theme Definition} & \textbf{Top 3 Code Patterns (Count, \%)} & \textbf{\# Codes (Total)} \\ \hline
Command Understanding, Analysis \& Generation & Centred on interpreting, analysing, or generating command-line inputs, including their structure, function, and potential threat relevance & Command Understanding/Analysis (861, 90.8\%), Command Threat Analysis (25, 2.6\%), Command Writing (17, 1.8\%) & 21 (948) \\ \hline

Text Generation \& Editing & Involves the generation, rewriting, or refinement of text, including grammar correction & Editing/Rewriting Text (130, 19.2\%), Grammar Correction (76, 11.2\%), Improving Incident Description (31, 4.6\%) & 116 (676) \\ \hline

Code, Script, and Regex Analysis & Involves writing, reading, or interpreting code, scripts, and regular expressions across various programming or automation contexts & Code Writing and Understanding (128, 37.5\%), Script Understanding (44, 12.9\%), Understanding ML concept (24, 7.0\%) & 41 (341) \\ \hline

Tool / Software / Application / System Understanding & Focused on gaining insight into how specific tools, software platforms, or system configurations work & Tool Understanding (54, 33.1\%), Software Understanding (14, 8.6\%), Registry Understanding (5, 3.1\%) & 62 (163) \\ \hline

LLM Chat: Greeting, Thanking, Version, Memory, Auditing & Centred around greetings, expressions of thanks, comments on versioning, or audit-related feedback & LLM Interaction (Thanking) (75, 48.1\%), LLM Interaction (Greeting) (20, 12.8\%), Casual Chat/Conversation (15, 9.6\%) & 25 (156) \\ \hline

File \& Malware Analysis & Related to understanding file attributes, locations, and content, often in the context of detecting malicious behaviour & File Understanding (94, 61.4\%), File Location Analysis (12, 7.8\%), File-Related Threat Analysis (6, 3.9\%) & 30 (153) \\ \hline

Attack, Security \& Threat Intelligence & Related to understanding attacks, malware types, vulnerabilities, and broader security threats for contextualisation or reporting & Attack Understanding (24, 17.0\%), Trojan Understanding (7, 5.0\%), Attack Type Understanding (6, 4.3\%) & 75 (141) \\ \hline

Query, Search, \& Data Processing & Involves the creation, refinement, or execution of search queries, often for use in log analysis, data filtering, or threat hunting & Query Writing (72, 60.0\%), Answering Multiple Choice Questions (Query) (8, 6.7\%), Query Understanding (6, 5.0\%) & 23 (120) \\ \hline

Network and Traffic Analysis & Focused on analysing IP addresses, URLs, and network traffic to understand communication patterns, identify threats, or interpret network behaviour & Network Traffic Understanding (18, 19.6\%), IP Address Analysis (14, 15.2\%), URL Analysis (8, 8.7\%) & 42 (92) \\ \hline

Access Control \& Authentication & Concerns remote access protocols, authentication methods, and policy or permission management & Policy \& Access Management (12, 13.0\%), Remote Management Protocol Understanding (10, 10.9\%), Authentication Method Understanding (9, 9.8\%) & 45 (92) \\ \hline

Email Security \& Analysis & Involving assessment of email rules, phishing detection, and analysis of potentially malicious email behaviours like forwarding, auto-deletion & Email Rule Understanding (21, 34.4\%), Email Rule Threat Assessment (13, 21.3\%), Email Rule Analysis (5, 8.2\%) & 16 (61) \\ \hline

Alert \& Log Analysis & Tasks focused on interpreting security alerts, analysing logs and event messages & Log Analysis (4, 11.8\%), Log Event Analysis (3, 8.8\%), Windows Event Management (3, 8.8\%) & 24 (34) \\ \hline

Others & Miscellaneous & Finding Movie Word Counts in a Range (5, 14.7\%), General Knowledge Related Task (4, 11.8\%), Keyword Research (SEO) (2, 5.9\%) & 21 (34) \\ \hline

Concept Definition & To define, clarify, or understand security-related concepts, systems, or identity mechanisms in depth & Concept Definition (7, 23.3\%), Term Understanding (6, 20.0\%), Concept Understanding (4, 13.3\%) & 15 (30) \\ \hline

Document/Evidence Search & Aimed at retrieving or summarising reports, templates, or locating relevant intelligence sources or documentation to support investigative work & Identifying Supporting Reports/Papers (13, 61.9\%), Obtaining a Report (4, 19.0\%), Fact Checking (1, 4.8\%) & 6 (21) \\ \hline

System Process Understanding \& Analysis & Involving analysis of system-level behaviours, understanding inter-process communication or OS-level services & OS Process Understanding (3, 21.4\%), Process Activity Investigation (3, 21.4\%), Process \& IPC Analysis (2, 14.3\%) & 8 (14) \\ \hline

Numerical Analysis & Involving math-related problem solving, percentage calculations, or interpreting numerically expressed patterns & Math Problem (6, 42.9\%), Numerical Analysis (3, 21.4\%), Math Problem Notation (1, 7.1\%) & 7 (14) \\ \hline

\end{tabular}
\end{table*}

\begin{table*}[htbp]
\centering
\caption{Summary of Subject Theme with Top 3 Codes}
\begin{tabular}{|p{5.70cm}|p{4.85cm}|p{4.5cm}|p{1.0cm}|}
\hline
\textbf{Theme} & \textbf{Theme Definition} & \textbf{Top 3 Code Patterns (Count, \%)} & \textbf{\# Codes (Total)} \\ \hline
command related (specific command, command line interfaces, multiple commands) & About command-line syntax, utilities, or individual commands used in operating systems or security-related tasks & command (831, 88.9\%), command - malicious (9, 1.0\%), command line utility (8, 0.9\%) & 60 (935) \\ \hline

message / paragraph / phrase / sentence / term / article / report / templates / formatting instructions & About written content such as phrases, sentences, messages, or templates, including wording, formatting, terminology, and stylistic adjustments & message/phrase/paragraph/sentence (277, 53.4\%), previous llm response (40, 7.7\%), term (21, 4.0\%) & 105 (519) \\ \hline

coding and scripting related (specific language, function, formula, code snippet, implementations, libraries, packages, development environment, errors, ecoding types, regex) & About code, scripts, or regular expressions, including syntax, structure, or functionality across various programming or scripting languages & script (41, 14.3\%), code (28, 9.8\%), coding error (10, 3.5\%) & 155 (286) \\ \hline

software \& tool related (different types,  OS, OS-level software, apps, applications, device drivers, development environments, installations, processes, OS-level processes, tools) & About specific software tools and applications, including features, setup options, or security-related configurations & tool (14, 6.2\%), software (10, 4.4\%), specific bitlocker configuration - bitlocker policy (9, 4.0\%) & 150 (225) \\ \hline

file and directory related (including naming conventions, types, file actions, permissions, file behaviours, sharing mechanism, directory, directories related) & About file names, file paths, directories, and file system locations relevant to detection or investigation & file (file name) (82, 52.9\%), file (file name) - specific location/folder (17, 11.0\%), directory (3, 1.9\%) & 51 (155) \\ \hline

LLM/GPT related (including privacy, chat aduditing, version, history) & Referencing gpt or large language models, including their versions, capabilities, or role-based configurations & llm\_gpt (119, 79.9\%), gpt version (9, 6.0\%), persona description (6, 4.0\%) & 15 (149) \\ \hline

attack \& security related (attack type, indicators, attacker motivations, concepts, solutions, scam, phishing, incidents, APT groups, threat groups, malware, malicious activity) & About malicious activity, including different types of malware (e.g., trojans, ransomware), attack techniques (e.g., phishing, brute force), and associated vulnerabilities that enable exploitation & malware type (12, 8.7\%), attack type (12, 8.7\%), network scanning method (4, 2.9\%) & 94 (138) \\ \hline

MDR \&  Query related (including edr, edr query, rules, ids, rules, configuration, log related) & About queries, filters, or rules related to managed detection and response (MDR) systems or log analysis platforms & query (20, 16.5\%), sumologic query (9, 7.4\%), Sumo logic (6, 5.0\%) & 70 (121) \\ \hline

alert, investigation, incident, security event related & About security alerts, incident details, investigation steps, or descriptions of suspicious or confirmed events & incident description (33, 30.8\%), alert description (29, 27.1\%), investigation description (13, 12.1\%) & 22 (107) \\ \hline

network communication mechanisms related (including sockets, http, ftp, ports, API, rdp, packet data, protocol, ip address, id address range, cidr) & About network-level elements such as IP addresses, port numbers, and remote access protocols like RDP & port number (7, 6.7\%), ip address (6, 5.7\%), regex - ip address (3, 2.9\%) & 77 (105) \\ \hline

email and mail flow management related (email rules, forwarding, mail flow config, client communication: emails, alerts, incidents, replies) & Different types of email rules and their interpretation & email (22, 26.8\%), email rule (11, 13.4\%), email rules (8, 9.8\%) & 38 (82) \\ \hline

Education and Interview related (training materials, Q\&A, interview questions) & Content focused on training, self-assessment, and preparation for interviews, such as scenario-based questions & scenario based interview questions (15, 20.5\%), mitre-related questions and answer (13, 17.8\%), different questions and detailed answers (10, 13.7\%) & 22 (73) \\ \hline

access management related (users, accounts, groups, roles, privileges, active directory, authentication) & Topics related to user identities, roles, groups, privileges, authentication methods, and account activity & authentication methods (2, 3.1\%), authentication method  - specific os (1, 1.5\%), organisation units - gpo - hosts (1, 1.5\%) & 64 (65) \\ \hline

general data types (numbers, strings, data, time, hash, hex, base64, table, lists, json) & Numbers, strings, data, time, hash, hex, base64, table, lists, JSON & decoding data (10, 17.2\%), math problem (6, 10.3\%), number division (3, 5.2\%) & 30 (58) \\ \hline

online sources and corroborating evidence or source related (research, papers, reports, online sources) & Evidence that could backup LLM or analyst & url (9, 30.0\%), corroborating evidence or source - research (2, 6.7\%), url click event (2, 6.7\%) & 19 (30) \\ \hline

general terms & Miscellaneous or off-topic queries & movie word counts (5, 23.8\%), world event (4, 19.0\%), joke (4, 19.0\%) & 10 (21) \\ \hline

device and hardware identify related (host, hostnames, domain names, storage device information) & Identifiers and information about hardware, hosts, domains, USB devices, or endpoints & domains (3, 25.0\%), usb details (2, 16.7\%), device (1, 8.3\%) & 9 (12) \\ \hline

organisation and agreement related (companies, competitors, organisational policies and agreements) & Organisational context, including companies, competitors, compliance requirements, policies & compliance/agreement (2, 22.2\%), company competitors (2, 22.2\%), internal organisation process (1, 11.1\%) & 7 (9) \\ \hline

\end{tabular}
\end{table*}

\end{document}